\documentclass[12pt]{elsarticle}
\biboptions{sort&compress}
\pdfoutput=1
\usepackage[numbers]{natbib}
\usepackage{url}
\usepackage[hmargin=2.35cm]{geometry}
\usepackage{caption}
\usepackage{subcaption}
\usepackage{xspace}
\usepackage{graphicx}
\usepackage{epstopdf}
\usepackage{mciteplus}
\usepackage{ifthen} 
\newboolean{uprightparticles}
\setboolean{uprightparticles}{false} 
\newboolean{articletitles}
\setboolean{articletitles}{true} 

\journal{Annual Review of Nuclear and Particle Science}

\begin{document}
\def\CP                {{\ensuremath{C\!P}}\xspace}
 \def\thebaroffset{0.18em}
\newcommand{\offsetoverline}[2][\thebaroffset]{\kern #1\overline{\kern -#1 #2}}%
\def\PD      {\ensuremath{D}\xspace}
\def\Dbar    {{\ensuremath{\offsetoverline{\PD}}}\xspace}
\def\D       {{\ensuremath{\PD}}\xspace}
\def\Db      {{\ensuremath{\Dbar}}\xspace}
\def\Dz      {{\ensuremath{\D^0}}\xspace}
\def\Dzb     {{\ensuremath{\Dbar{}^0}}\xspace}
\def\DzDzb   {{\ensuremath{\D^0\Dbar{}^0}}\xspace}

\title{{\bf Mixing and $C\!P$ violation in the charm system}}

\author[siegen]{A.~Lenz} 
\author[oxford]{G.~Wilkinson}

\address[siegen]{Theoretische Physik 1, Fakult{\"a}t IV, Department Physik, Universit{\"a}t Siegen, Siegen,  D-57068, Germany}
\address[oxford]{University of Oxford, Denys Wilkinson Building, Keble Road,  OX1 3RH, United Kingdom}

\markboth{Lenz, Wilkinson}{Mixing and CPV in the charm system}

\begin{abstract}
In recent years charm physics has undergone a renaissance, one which has been catalysed by an unexpected and impressive set of experimental results from the $B$-factories, the Tevatron and LHCb.  The existence of $\Dz\Dzb$ oscillations is now well established, and the recent discovery of $\CP$ violation in $\Dz$ decays has further renewed interest in the charm sector.  In this article we review the current status of charm-mixing and $\CP$-violation measurements, and assess their agreement with theoretical predictions within the Standard Model and beyond. We look forward to the great improvements in experimental precision that can be expected over the coming two decades, and the prospects for corresponding advances in theoretical understanding.
\end{abstract}

\endfrontmatter

\tableofcontents

\section{Introduction}
\label{sec:intro}

Charm played a pivotal role in the construction of the Standard Model (SM). Its existence was invoked in the discovery of the GIM mechanism~\cite{PhysRevD.2.1285}, and it was the observation of the $J/\psi$ in 1974 that convinced the community of the validity of the quark model~\cite{PhysRevLett.33.1404,PhysRevLett.33.1406}.  It was soon appreciated that mixing and $\CP$-violating phenomena are expected to manifest themselves at a considerably lower level in charm than in the beauty system, and indeed for many years all searches for such effects were frustrated.  The last fifteen years, however, has seen enormous experimental progress, a trajectory that looks set to continue over the coming two decades.

In this review we summarise the experimental and theoretical status of charm-mixing and $\CP$-violation studies.  We describe the key measurements that have established our current understanding, and look forward to the advances that can be expected in the coming years.  We discuss the theoretical tools that have been developed to confront the data, and pay particular attention to the  challenges that charm studies bring in this respect.  We stress the great potential of charm for revealing signs of physics beyond the Standard Model, and assess some of the most interesting existing experimental results in this context.


\section{Theoretical framework}
\label{sec:framework}

In this review we shall consider $\Dz\Dzb$ mixing and direct $\CP$ violation in singly Cabibbo-suppressed (SCS) decays $c \to s \bar{s} u$ 
({\it e.g.} $\Dz \to K^+ K^-$) and 
$c \to d \bar{d} u$ ({\it e.g.} $\Dz \to \pi^+ \pi^-$). 
SCS decays have a penguin
contribution in addition to the tree-level
contribution, leading to the possibility of $\CP$-violating effects, in contrast to
the case of Cabibbo-favoured (CF) decays  
$c \to s \bar{d} u$ ({\it e.g.} $\Dz \to \pi^+ K^-$) and  doubly Cabibbo-suppressed (DCS) decays  $c \to d \bar{s} u$ 
({\it e.g.} $\Dz \to K^+ \pi^-$),
which cannot have penguin contributions.

In order to sum up large logarithms
of the form $\alpha_s(m_c)^m \ln^n (m_c^2/M_W^2)$ ($m \geq n$) to all orders, weak decays of charmed hadrons are treated
via the effective Hamiltonian
 ${\cal H}_{\rm eff}$, where all heavy (that is heavier than the charm quark) degrees-of-freedom have been integrated out (see {\it e.g.} Ref.~\cite{Buras:2020xsm} for more details). Here
 \begin{eqnarray}
 {\cal H}_{\rm eff} & = & \frac{G_F}{\sqrt{2}}
 \left\{ \lambda_d \left[ C_1(\mu) Q_1^d  + C_2(\mu) Q_2^d \right]
 +\lambda_s \left[ C_1(\mu) Q_1^s  + C_2(\mu) Q_2^s \right]
 + \lambda_b \sum \limits_{i \geq 3} C_i(\mu) Q_i
 \right\} 
\, , \nonumber
\\
 \end{eqnarray}
 with the Cabibbo-Kobyashi-Maskawa (CKM)~\cite{Cabibbo:1963yz,Kobayashi:1973fv} structures $\lambda_x = V_{cx} V^*_{ux}$ .
 The Wilson coefficients $C_i(\mu)$ have been calculated in perturbation theory (see {\it e.g.} Ref.~\cite{Buchalla:1995vs}).
 $Q_1$ and $Q_2$ denote tree-level four-quark operators
 \begin{eqnarray}
   4  Q_1^q  =    \bar{q}^\alpha \gamma_\mu (1- \gamma_5) c^\beta  \times  \bar{u}^\beta \gamma^\mu (1- \gamma_5) q^\alpha   \, , 
     &&
     4 Q_2^q  =   \bar{q}^\alpha \gamma_\mu (1- \gamma_5) c^\alpha \times \bar{u}^\beta \gamma^\mu (1- \gamma_5) q^\beta  \, , 
 \end{eqnarray}
 where $\alpha$ and $\beta$ are colour indices. 
 $Q_{i \geq 3}$ denote penguin operators, where the heavy $b$-quark and the $W$-boson have been integrated out. 
 The unphysical renormalisation scale dependence of the Wilson coefficients cancels order-by-order with the $\mu$-dependence of the matrix elements of the four-quark operators.
 Amplitudes describing the decay of a neutral $D$ meson into a final state $f$ will be  denoted by
 \begin{equation}
     A_f = \langle f| {\cal H}_{\rm eff} |D^0 \rangle
     ; \, \,
    \bar{A}_f = \langle f| {\cal H}_{\rm eff} |\bar{D}^0 \rangle
     ; \, \,
      A_{\bar{f}} = \langle \bar{f}| {\cal H}_{\rm eff} |D^0 \rangle
      ; \, \,
     \bar{A}_{\bar{f}} = \langle \bar{f}| {\cal H}_{\rm eff} |\bar{D}^0 \rangle.
    \end{equation}
 
 Weak decays and mixing of hadrons containing a
 charm quark show several peculiarities compared
 to decays of $b$ hadrons, which make their
 theoretical description considerably more
 challenging.
First, the corresponding CKM elements  are to a large extent real and therefore do not leave  much space for $\CP$-violating effects. According to the web-update of Ref.~\cite{Charles:2004jd} (for similar results see Ref.~\cite{Bona:2006ah}) we find for the central values of the CKM elements involving the charm quark
\begin{eqnarray}
 V_{\rm cd}  =  -0.2245 - 2.6 \cdot 10^{-5} I \, , 
 \hspace{0.5cm}
 V_{\rm cs}  =  0.97359 - 5.9 \cdot 10^{-6} I \, ,  \, \, \,
&&
 V_{\rm cb}  =   0.0416   \, .
 \end{eqnarray}
These numbers were determined assuming the validity of the SM and in particular the unitarity of the CKM matrix and they 
might change in the presence of beyond-the-SM (BSM) effects.
Moreover, the value
of a single matrix element is unphysical and depends on the phase convention, which was chosen here such that $V_{cb}$ is real.
In $D$-mixing and SCS  $D$-meson decays the physical combinations
$\lambda_q = V_{cq} V^*_{uq}$ will arise. We find as central values
\begin{eqnarray}
\lambda_d =  - 0.21874           - 2.51 \cdot 10^{-5} I \, , &&
\\
\lambda_s =  + 0.21890           - 0.13 \cdot 10^{-5} I 
\, ,
&& \, \, \, \, \, \, \, \, \, \, \, \,
\lambda_b =  -1.5 \cdot 10^{-4} + 2.64 \cdot 10^{-5} I
\, .
\end{eqnarray}
Again we observe for $\lambda_d$ and $\lambda_s$ only  negligible imaginary parts, and that their real parts 
are of almost exactly the same size but of opposite sign. Thus the unitarity relation 
$\lambda_d + \lambda_s + \lambda_b =0$ is fulfilled to a good approximation for the first two
generations alone, {\it i.e.} $\lambda_d + \lambda_s \approx 0$.
This is one of the two reasons why 
the GIM mechanism \cite{PhysRevD.2.1285} can be extremely pronounced in the charm sector. 
The modulus of $\lambda_b$ is much smaller than $|\lambda_{d,s}|$,
but its imaginary part has a similar size as its real one. Therefore this latter CKM combination could be important for 
the potential size of  $\CP$ violation in the charm system within the SM. 

\begin{figure}[htb]
   \centering
\begin{subfigure}{.36\textwidth}
\centering
\includegraphics[width=0.95\textwidth]{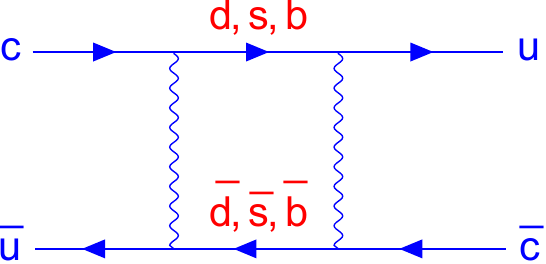}
\end{subfigure}
\begin{subfigure}{.36\textwidth}
  \centering
\includegraphics[width=0.95\textwidth]{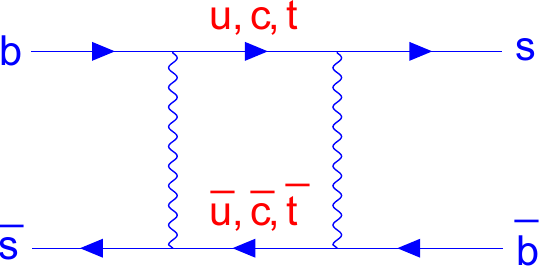}
\end{subfigure}
\\
\begin{subfigure}{.36\textwidth}
\centering
\centering\includegraphics[width=0.95\textwidth]{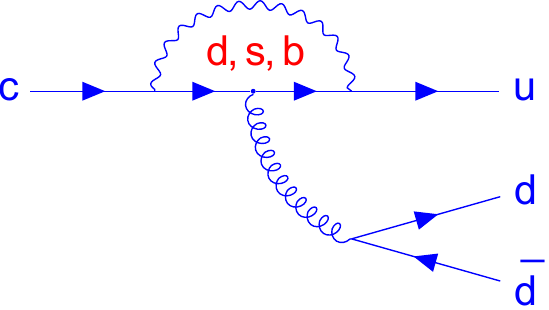}
\end{subfigure}
\begin{subfigure}{.36\textwidth}
\centering
\includegraphics[width=0.95\textwidth]{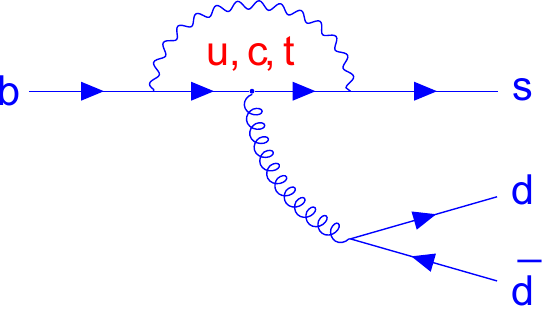}
\end{subfigure}
\caption{Internal quarks in loop-induced transitions (neutral-meson mixing in top line and penguin decays in lower line) of $D$ and $B$ mesons.}
        \label{fig:loop}
\end{figure}

Next, we find that in loop contributions arising in $D$-mixing or in penguin contributions, the charm quark can 
transform into an internal $d$, $s$ or $b$ quark, as indicated in  Fig.~\ref{fig:loop}. In the case of $b$-hadron decays the corresponding internal 
quarks would be $u$, $c$ and $t$. 
The loop functions depend on the square of the virtual quark mass divided by the $W$-boson mass. We thus find  the following size of  loop-function parameters in the charm and the beauty sector:
\begin{eqnarray}
\left( \frac{m_d}{M_W} \right)^2 \approx 0 \, , \hspace{1.1cm}   &&
\left( \frac{m_u}{M_W} \right)^2
\approx 0 \, ,
\\
\left( \frac{m_s}{M_W} \right)^2 \approx 1.3 \cdot 10^{-6} \, ,&&
\left( \frac{m_c}{M_W} \right)^2
\approx 2.5 \cdot 10^{-4} \, ,
\\
\left( \frac{m_b}{M_W} \right)^2 \approx 2.8 \cdot 10^{-3} \, ,&&
\left( \frac{m_t}{M_W} \right)^2
\approx 4.5 \,.
\end{eqnarray}
Hence in the charm sector all internal masses are very close to zero and therefore largely identical - for identical 
quark masses the GIM mechanism gives a zero result. This is the second reason for the severeness of the GIM mechanism
in the charm sector, in contrast to the $b$ system, where we have the large value of the top quark mass, strongly 
breaking the GIM mechanism.

A third peculiarity lies in the  fact that the value of the charm-quark mass is close to the hadronic scale 
$\Lambda_{QCD}$, which makes the applicability of a Taylor expansion in $\Lambda_{QCD}/m_c$  a priori questionable. 
This uncertainty should, however, be resolved by an as precise as possible determination of the expansion parameter
$\langle Q \rangle / m_c$ with theoretical state-of-the-art methods and not by  simple dimensional analysis.
 Observables that are not affected by severe GIM cancellations, for example lifetimes of charmed
hadrons, are of course much better suited than {\it e.g.} $D$-mixing for studying this issue. 

Finally the value of the strong coupling at the charm scale, $\alpha_s(m_c)$ is large. At 
 5-loop accuracy at the charm pole mass scale ($m_c = 1.67$\,GeV) we have
\begin{equation}
  \alpha_s(m_c)  = 0.33 \pm 0.01 \, .
\end{equation}
This large value clearly points towards the important effect of higher-order perturbative contributions, and maybe even to the existence of sizeable  non-perturbative effects.


\section{Experimental overview}
\label{sec:exp-overview}

Almost all measurements on \CP-violation and mixing discussed in this review were performed either at BaBar and Belle, the $e^+e^-$ $B$-factory experiments that operated in the first decade of the millennium, or at CDF and LHCb, hadron-collider experiments situated at the Tevatron and LHC, respectively.  Table~\ref{tab:yields} shows the signal yields at these experiments for three example decay modes.

\begin{table}[htb]
\caption{Reported and expected signal yields in example decay modes, flavour-tagged through the process $D^{\ast +} \to \Dz \pi^+$, and including semileptonic tags for the LHCb yields in $\Dz \to K^0_{\rm S}\pi^+\pi^-$. The $\Dz \to \pi^+\pi^-\pi^0$ expectations are scaled from the existing measurements according to the ratio of integrated luminosities and, for the LHCb Upgrades, accounting for a change in cross-section with collision energy. A factor two improvement in trigger efficiency is also assumed for LHCb U\,I and U\,II.
}\label{tab:yields}
    \centering
    \begin{tabular}{l rrr rrr rrr}\hline
                & \multicolumn{3}{c}{$\Dz \to K^+\pi^-$} &   \multicolumn{3}{c}{$\Dz \to \pi^+\pi^-\pi^0$} &  
                   \multicolumn{3}{c}{$\Dz \to K^0_{\rm S}\pi^+\pi^-$} \\ \hline
    BaBar/Belle & 11.5k & $1.0\,{\rm ab}^{-1}$ & \cite{Ko:2014qvu}
                & 126k  & $0.5\,{\rm ab}^{-1}$ & 
                \cite{TheBABAR:2016gom}
                & 1.2M  & $0.9\,{\rm ab}^{-1}$ & \cite{Peng:2014oda} \\
    CDF         & 32.7k   &  $9.6\,{\rm fb}^{-1}$ &  \cite{Aaltonen:2013pja}
                & \multicolumn{3}{c}{/}
                & 0.3M  &  $6.0\,{\rm fb}^{-1}$ & \cite{Aaltonen:2012nd} \\
    LHCb        & 722k  &  $5.0\,{\rm fb}^{-1}$ & \cite{Aaij:2017urz}
                & 566k   &  $2.0\,{\rm fb}^{-1}$  & \cite{Aaij:2014afa}
                & 2.3M  &  $3.0\,{\rm fb}^{-1}$  & \cite{Aaij:2019jot} \\
                & &&& &&& && \\
    Belle II    & 225k & $50\,{\rm ab}^{-1}$ & \cite{Kou_2019}
                & 13M & $50\,{\rm ab}^{-1}$ & 
                & 67M & $50\,{\rm ab}^{-1}$ & \cite{Kou_2019} \\
    LHCb U\,I & 25M & $50\,{\rm fb}^{-1}$  & \cite{lhcbupgrade2}
                & 44M & $50\,{\rm fb}^{-1}$  &
                & 598M & $50\,{\rm fb}^{-1}$  & \cite{lhcbupgrade2}\\
    LHCb U\,II & 170M & $300\,{\rm fb}^{-1}$  & \cite{lhcbupgrade2}
                & 291M & $300\,{\rm fb}^{-1}$  &
                & 3990M & $300\,{\rm fb}^{-1}$  & \cite{lhcbupgrade2}\\ \hline
    \end{tabular}
\end{table}

At the $B$-factories charm hadrons are studied through their continuum production.  The process $e^+e^- \to \gamma^\ast \to c\bar{c}$ has a cross-section of around 1.3\,nb at $\sqrt{s} = 10.58\,{\rm GeV}$, which exceeds that of $e^+e^- \to \Upsilon(4S)$.  The low-multiplicity environment allows the efficient reconstruction of many decay modes, including those involving $\pi^0$ mesons. In measurements that require  knowledge of the initial flavour of the neutral charm meson of interest, this information is obtained by {\it flavour tagging} this meson  through the strong decay $D^{\ast +} \to D^0 \pi^+$.  Here the accompanying particle is named the {\it slow pion} on account of its low momentum in the $D^{\ast +}$ rest frame.  Its charge distinguishes between $\Dz$ and $\Dzb$ mesons.

At hadron machines the most prolific source of charm is through {\it prompt production} in the initial $pp$ (or $p\bar{p}$) collision.  Here the  cross-section is enormous; for example $\sigma(pp \to c\bar{c}X) = (2369 \pm 192)\,{\rm \mu}b$ at $\sqrt{s}=13\,{\rm TeV}$ within the approximate acceptance of LHCb~\cite{Aaij:2015bpa}.
The main challenge in exploiting these events lies in the triggering, as most of the decays of interest are fully hadronic, and the decay products have relatively low transverse momentum.  
Only experiments with dedicated heavy-flavour triggers, such as LHCb or CDF,  have been successful in accumulating large samples through this source. As at the $B$-factories, neutral mesons may be flavour-tagged through $D^{\ast +}$ decays.

Analyses have also been performed at LHCb that make use of {\it secondary production}, where the charmed particle arises from the semileptonic decay of a $b$ hadron.  Although the cross-section for beauty production is around twenty times lower than that of charm, the events are relatively easy to select because of the high efficiency for triggering semileptonic $b$-hadron decays.  The charge of the muon also provides a  flavour tag.  A final benefit of such a sample is that the selection efficiency is relatively uniform with decay time, in contrast to a prompt sample where the trigger requirements tend to reject decays with low lifetimes.  It should also be noted that secondary-charm production (not necessarily in semileptonic decays) contaminates the prompt samples, typically at the percent level, and growing with decay time.  In time-dependent measurements  this contamination induces a bias, as the birth coordinate of the secondary $\Dz$ meson is wrongly assigned to the primary vertex associated with the $pp$ collision.

Studies performed at $e^+e^-$ machines at and just above charm threshold make an important contribution to our knowledge of charm physics.  The CLEO-c experiment collected $0.8\,{\rm fb}^{-1}$ of data at the $\psi(3770)$ resonance, and BESIII has so far accumulated $2.9\,{\rm fb}^{-1}$, with the expectation of increasing this total to around $20\,{\rm fb}^{-1}$~\cite{Ablikim:2019hff}.  Although the sample sizes are not large enough to permit competitive \CP-violation and mixing studies, the very clean environment and lack of fragmentation particles permit unique measurements of branching ratios, $\CP$-conserving strong phases  and associated parameters, that are important inputs to analyses performed at other facilities. Many of the results are achieved through a double-tag technique, in which both charm mesons are reconstructed and where measurements can benefit from the quantum coherence of the final state ({\it e.g.} see Ref.~\cite{Ablikim:2020lpk}).

Over the coming decade results of ever increasing sensitivity are expected from LHCb, which is currently undergoing an upgrade (Upgrade~I or U~I) that will allow for operation at higher luminosity. The goal is to increase the integrated luminosity of the experiment from the $3\,{\rm fb}^{-1}$ (Run 1 at 7 and 8\,TeV) plus $6\,{\rm fb}^{-1}$ (Run 2 at 13\,TeV) currently on tape to a total of $50\,{\rm fb}^{-1}$. A full software trigger will be deployed that will significantly improve efficiency for many decays, particularly in charm physics.
The LHCb results will be complemented by measurements from Belle II, which is now data taking and aims to accumulate around $50\,{\rm ab^{-1}}$ of integrated luminosity~\cite{Kou_2019}.  Plans are evolving for an Upgrade~II (U~II) of LHCb, which will operate in the next decade and allow the experiment to collect a total data set of $300\,{\rm fb}^{-1}$~\cite{lhcbupgrade2}.  
Table~\ref{tab:yields} includes the predicted signal yields for these projects that illustrate well the enormous increase in statistical precision that can be expected. 
If approved, a Super Tau Charm Factory  will run in the 2030s in China, with the potential of integrating ${\cal{O}}(1\,{\rm ab}^{-1})$ at the $\psi(3770)$ resonance~\cite{Luo:2019xqt}. A similar facility is also proposed to be located in Novosibirsk, Russia~\cite{EIDELMAN2015238}.

Other experiments are being planned of very different natures to those discussed above, which would also be suitable for performing precise studies in charm physics. 
The FCC-ee, scheduled to operate in the 2040s, would produce around $6 \times 10^{11}$ $c\bar{c}$ pairs from $Z^0$ decays in a relatively clean analysis environment~\cite{Abada:2019lih}. CEPC in China would have similar capabilities. 
Ideas exist to allow a high intensity proton beam, such as that proposed for the Beam Dump Facility at CERN, to impinge on a system of tungsten targets, which would then  produce a comparable number of charm hadrons to that  at LHCb Upgrade II (see discussion of the TauFV experiment in Refs.~\cite{Ahdida:2650896,group2019physics}).


\section{The $\DzDzb$ system: mixing and $\CP$ violation in mixing}
\label{sec:mixing}

\subsection{Mixing formalism for neutral $D$ mesons}
\noindent
The weak eigenstates of the neutral $D$ mesons ($\Dz = (c \bar{u})$ and $\Dzb = (\bar{c} u)$ with the convention $\CP | \Dz \rangle = -  | \Dzb \rangle$ and $\CP | \Dzb \rangle = -  | \Dz \rangle$, which is assumed throughout this review) mix under the weak interaction and the time evolution of the neutral $D$ mesons is governed by the mass eigenstates $D_{1,2} = p \Dz \mp q \Dzb$.
Diagonalisation of the 2$\times$2 matrix describing this mixing 
gives the following  eigenvalue equations:
\begin{eqnarray}
\Delta M_D^2 - \frac{\Delta \Gamma_D^2}{4} 
=  
4 \left| M_{12}^D \right|^2 - \left| \Gamma_{12}^D \right|^2 =: a\, ,
&&
\Delta M_D \Delta \Gamma_D 
 =  
4 \left| M_{12}^D \right| \left| \Gamma_{12}^D \right| \cos (\phi_{12}^D) =: b\, ,
\end{eqnarray}
with the mass difference $\Delta M_D = M_1 - M_2$ and the decay rate difference $ \Delta \Gamma_D = \Gamma_1 - \Gamma_2$ of the mass 
eigenstates of the neutral $D$ mesons. The mass difference is by definition positive, while the decay rate difference can have any sign - below we will only specify the positive one. The box diagrams giving rise to $D$ mixing can have internal
$d$, $s$ and $b$ quarks, to be compared to $u$, $c$, $t$ in the $B$ sector (see Fig.~\ref{fig:loop}). 
$M_{12}^D$ denotes the 
dispersive part of the box diagram, $\Gamma_{12}^D$ the absorptive part, and the relative phase 
of the two is given by $\phi_{12}^D =  \arg (-M_{12}^D / \Gamma_{12}^D)$. Unlike in the $B$ 
system, where $|\Gamma_{12}/M_{12}| \ll 1$ holds, the expressions for  $\Delta M_D$ and 
$\Delta \Gamma_D$ in terms of $M_{12}^D$ and $\Gamma_{12}^D$ cannot be simplified,  
and to know $\Delta M_D$ both $M_{12}^D$ and $\Gamma_{12}^D$ have to be determined. In general we get
\begin{eqnarray}
\Delta \Gamma_D  =  \sqrt{2 \left( \sqrt{a^2 + b^2} -a \right)} \, ,
& \, \, \, \, &
\Delta M_D  =  \sqrt{\frac{\sqrt{a^2 + b^2} +a}{2} } \, .
\end{eqnarray}
On the other hand it is well known that bounds like $|\Delta \Gamma_D| \leq 2 |\Gamma_{12}^D |$ and $\Delta M_D \leq 2 |M_{12}^D |$ 
hold generally for any values of $M_{12}^D$ and $\Gamma_{12}^D$ \cite{Nierste:2009wg,Jubb:2016mvq}.
Experimental studies of charm mixing are sensitive to the parameters 
\begin{eqnarray}
x \equiv \frac{\Delta M_D}{ \Gamma_D}  \, ,
&&
y \equiv \frac{\Delta \Gamma_D}{2 \Gamma_D}  \, ,
\end{eqnarray}
where $\Gamma_D$ denotes the total decay rate of the neutral $D$ mesons. Measurements, whose combined results are summarised later in Eq.~\ref{eq:xyvalues}, show $y \sim 1\%$, and clearly establish the existence of a finite width difference in the neutral charm system. Current indications are that $x$ has a similar value, although the possibility of 
having a vanishing mass difference is still not excluded.

However, if one assumes a small $\CP$-violating phase $\phi_{12}^D$ one obtains
\begin{eqnarray}
\Delta \Gamma_D & = & 2 |\Gamma_{12}^D| \left\{
1 - \frac{4}{4+r} \frac{(\phi_{12}^D)^2}{2} + {\cal O} \left[(\phi_{12}^D)^4\right]\right\} \, ,
\label{DG_small_phi}
\\
\Delta M_D & = & 2 |M_{12}^D| \left\{
1 - \frac{r}{4+r} \frac{(\phi_{12}^D)^2}{2} + {\cal O} \left[(\phi_{12}^D)^4 \right]\right\} \, ,
\label{DM_small_phi}
\end{eqnarray}
with $r= |\Gamma_{12}^D|^2/|M_{12}^D|^2 $. Note that the ratios $4/(4+r)$ and $r/(4+r)$ can only vary between zero and one, for
any value of $r$. Thus for values of up to $\phi_{12}^D \approx 0.45 \approx 25^\circ$ the corrections due to the weak phase in Eq.~(\ref{DG_small_phi}) and Eq.~(\ref{DM_small_phi})
are less than 10\% and therefore clearly small compared to the theoretical precision.
Using the approximations from Eq.~(\ref{DG_small_phi}) and Eq.~(\ref{DM_small_phi}) experiment tells us 
$r \approx 4y^2/x^2 \approx (13.5^{+16}_{-6}) $. This is in stark contrast to the $B$-system, where we have  
$r  \approx 2 \cdot 10^{-5} $. 

In the literature  "theoretical"
mixing parameters are commonly defined as
\begin{equation} 
x_{12} = 2 \frac{|M_{12}^D|}{\Gamma_D} \, , \hspace{1cm} 
y_{12} =   \frac{|\Gamma_{12}^D|}{\Gamma_D} \, ,  
\end{equation}
which coincide with $x$ and $y$ up to corrections of order $(\phi_{12}^D)^2$.

Time-dependent $\CP$ asymmetries of neutral $D$-meson decays into a $\CP$
eigenstate $f$ ({\it e.g.} $K^+ K^-$, $\pi^+ \pi^-$) with eigenvalue $\eta_{\CP}$ depend on the quantity
\begin{eqnarray}
\lambda_f & = & \frac{q}{p} \frac{\bar{A}_f}{A_f} = - \eta_{\CP}
\left| \frac{q}{p}  \right| e^{i \phi^D} \,
\end{eqnarray}
where penguin corrections have been neglected (see {\it e.g.} Ref.~\cite{Kagan:2009gb}). Here
$\phi^D$ is the relative weak phase between the
mixing contribution and the decay amplitudes.
In the case of non $\CP$-eigenstates an additional strong
phase will arise. Neglecting subleading CKM
structures the authors of Ref.~\cite{Kagan:2009gb} find a relation between $\phi^D$ and $\phi_{12}^D$:
\begin{eqnarray}
\tan 2 \phi^D =  - \frac{\sin 2 \phi_{12}^D}{\cos 2 \phi_{12}^D + \frac{r}{4}} \rightarrow
\phi^D \approx 
 -  \frac{4}{4 + r} \phi_{12}^D
 +{\cal O} \left[(\phi_{12}^D)^3 \right]
 \approx
(0.2 \pm 0.1) \, \phi_{12}^D \, ,
\end{eqnarray}
where we have expanded 
in small values of $\phi_{12}$ and
inserted the experimental central
values of $x$ and $y$.
The ratio $q/p$ is also
obtained by the diagonalisation of the
mixing matrix \footnote{Ref.~\cite{Kagan:2020vri} uses a different convention and thus obtains different signs in the expansion in $\phi_{12}$.}
\begin{equation}
    \frac{q}{p} = - \frac{2 (M_{12}^D)^* - i (\Gamma_{12}^D)^*}{\Delta M_D + \frac{i}{2} \Delta \Gamma_D} \, \, \,  \mbox{with} \, \, \,
    \left| \frac{q}{p}  \right|  = 
1 - \frac{2 \sqrt{r}}{4 + r} \phi_{12}^D + \frac{2r}{(4+r)^2} (\phi_{12}^D)^2 + {\cal O} \left[(\phi_{12}^D)^3 \right] \, . 
\end{equation}

In the experimental discussion below we will
define the quantities $y_{CP}$ 
(see Eq.~(\ref{eq:ycp_def}) and
Ref.~\cite{Bergmann:2000id}) and 
$A_\Gamma$ (see Eq.~(\ref{eq:ACP}) 
and Ref.~\cite{Nir:2005js}); neglecting 
direct $\CP$ violation and expressing them
in terms of $\phi^D$ and $p/q$ they read
\begin{eqnarray}
y_{CP} & = & \frac{y}{2} \left( \left| \frac{q}{p} \right|+ \left| \frac{p}{q} \right|\right) \cos \phi^D -  \frac{x}{2} \left( \left| \frac{q}{p} \right|- \left|\frac{p}{q} \right|\right)  \sin \phi^D \, \label{eq:ydef1}
\\
A_\Gamma & = & \frac{y}{2} \left( \left| \frac{q}{p} \right|- \left|\frac{p}{q} \right|\right)  \cos \phi^D -  \frac{x}{2} \left( \left| \frac{q}{p} \right|+ \left|\frac{p}{q} \right|\right) \sin \phi^D \, .
\label{eq:agdef1}
\end{eqnarray}
Expanding in the mixing phase $\phi_{12}^D$ we find
\begin{eqnarray}
y_{CP} & = &  y_{12} \left\{ 1 - \frac{32 (\phi_{12}^D)^2}{ (4+r)^2} 
+{\cal O} \left[(\phi_{12}^D)^3 \right] \right\}
\approx  y_{12} \left\{ 1 - (0.1\pm 0.1) \, \phi_{12}^2 +{\cal O} \left[ (\phi_{12}^D)^3 \right] \right\}
\, ,
\\
A_\Gamma & = &  2   \frac{2x_{12} -\sqrt{r}y_{12}}{4+r}  \phi_{12}^D + {\cal O} \left[(\phi_{12}^D)^3 \right]
\approx
  - 2 \cdot 10^{-3}  \phi_{12}^D + {\cal O} \left[(\phi_{12}^D)^3 \right]
\, ,
\label{AGamma_Taylor}
\end{eqnarray}
where we have  inserted on the r.h.s. of the last two equations the estimated value of $r$.

$\Gamma_{12}^D$ can be expressed in terms of on-shell box diagrams differing in the 
internal quarks - $(s \bar{s})$, $(s \bar{d})$, $(d \bar{s})$ and $(d \bar{d})$, (again see Fig.~\ref{fig:loop}).
Using CKM unitarity ($\lambda_d + \lambda_s +\lambda_b = 0$)  one gets
\begin{eqnarray}
\Gamma_{12}^D & = & -   \lambda_s^2 \left(\Gamma_{ss}^D - 2 \Gamma_{sd}^D + \Gamma_{dd}^D\right)
                    + 2 \lambda_s \lambda_b \left(\Gamma_{sd}^D - \Gamma_{dd}^D \right)
                    -   \lambda_b^2 \Gamma_{dd}^D \, .
\label{Gamma12D}
\end{eqnarray}
Eq.~(\ref{Gamma12D}) shows a very pronounced CKM hierarchy \cite{Charles:2004jd,Bona:2006ah}): 
\begin{eqnarray}
\lambda_s^2  & = &  +4.79 \cdot 10^{-2} - 5.85 \cdot 10^{-7} I \, ,
\\
2 \lambda_s \lambda_b  & = & 
- 6.61 \cdot 10^{-5} 
+ 1.16 \cdot 10^{-5} I \, ,
\\
\lambda_b^2  &= & + 2.21 \cdot 10^{-8}  - 7.99 \cdot 10^{-9}  I \, .
\end{eqnarray}
The CKM suppression of the second term in Eq.~(\ref{Gamma12D}) relative to the first one is roughly three orders of magnitude. A similar suppression is found for the third term compared to the second one.
It is interesting to note that the second term has a significant phase. Moreover in the exact $SU(3)_F$ limit $\Gamma_{ss}^D = \Gamma_{sd}^D = \Gamma_{dd}^D$ holds and 
the first two terms  vanish and only the tiny contribution from 
the third term survives. 
Below we will discuss in more detail the GIM cancellations taking place in the first two terms of this expression.
The determination of $M_{12}^D$ involves in addition box diagrams with internal $b$ 
quarks; in contrast to $\Gamma_{12}^D$ the dispersive part of the diagrams has 
now to be determined. Denoting the dispersive part of a box diagram with internal $x$ and $y$ 
quarks by $M_{xy}^D$ and using again CKM unitarity one gets the following structure:
\begin{eqnarray}
M_{12}^D \! & \! = \! &  \! \! \! \lambda_s^2          \left[ M_{ss}^D \! - 2 M_{sd}^D + M_{dd}^D \right]
              \! + 2 \lambda_s \lambda_b  \left[ M_{bs}^D \! -   M_{bd}^D \! - M_{sd}^D + M_{dd}^D \right]
              \!+   \lambda_b^2          \left[ M_{bb}^D \! - 2 M_{bd}^D + M_{dd}^D \right]\, .
\label{M12D}
\nonumber
\\
\end{eqnarray} 
In the case of neutral $B$ mesons the third term (replacing $b,s,d \to t,c,u$) is clearly dominant, while for charm the extreme CKM suppression of $\lambda_b$ might be compensated by a less pronounced 
GIM cancellation \cite{PhysRevD.2.1285}, and so all three terms of 
Eq.~(\ref{M12D}) must be considered.

\subsection{Theoretical approaches}
\noindent 
For the theoretical determination of $M_{12}^D$ and $\Gamma_{12}^D$ one can use a 
quark-level (inclusive) or a
hadron-level (exclusive) description.
We will briefly review the status of both approaches.

\subsubsection{Inclusive approach} \noindent
The inclusive approach for $\Gamma_{12}^D$ is based on the heavy-quark expansion (HQE) 
\cite{Khoze:1983yp,Shifman:1984wx,Bigi:1991ir,Bigi:1992su,Blok:1992hw,Blok:1992he,Beneke:1998sy}, which 
works very well for the $B$ system \cite{Lenz:2014jha,Artuso:2015swg,Kirk:2017juj}.
Interestingly the HQE can successfully explain the large lifetime ratio $\tau(D^+) / \tau (D^0)$  
\cite{Lenz:2013aua,Kirk:2017juj}, indicating an expansion parameter $\Lambda/m_c \approx 0.3$. 
Lifetimes have the advantages of being free from GIM cancellations and, to make statements concerning 
the convergence of the HQE in the charm system more sound, future experimental and theoretical 
studies of other charmed mesons and charmed baryon lifetimes will be necessary.

Considering only a single diagram contributing to $\Gamma_{12}^D$, {\it e.g.} with only internal
$s \bar{s}$ quarks, the HQE predicts 
(the non-perturbative matrix elements of dimension-six operators have been
determined in Refs.~\cite{Carrasco:2014uya,Carrasco:2015pra,Bazavov:2017weg,Kirk:2017juj})
a value for y that is
five times larger than the experimental result~\cite{Lenz:2016fcv}.
However, if one 
applies  HQE to the whole expression of Eq.~(\ref{Gamma12D}) 
one encounters huge GIM suppression and obtains a result lying about four to five orders 
of magnitude below experiment
\cite{Lenz:2020efu}! 

Taking account of the success of the HQE in the $B$ system (where the 
expansion parameter is only a factor of three smaller) and for $D$-meson lifetimes 
it appears unlikely that HQE simply fails by four orders of magnitude in charm mixing. Instead the problem seems
to be rooted in the severe GIM cancellations.
There have been several solutions suggested in the literature for this puzzle:
the first idea, studied in  
Refs.~\cite{Georgi:1992as,Ohl:1992sr,Bigi:2000wn,Bobrowski:2010xg},
is a lifting of the  GIM cancellation in the first and second term of Eq.~(\ref{Gamma12D})
by higher terms in the HQE, thereby overcompensating the $\Lambda/m_c$ suppression. First estimates 
of the dimension-nine contribution in the HQE for $D$-mixing~\cite{Bobrowski:2012jf}
indicate an enhancement compared to the leading dimension-six terms, but one unfortunately not 
large enough to explain the experimental value. Here a full theory determination of the HQE
terms of dimension nine and twelve will provide further insight.
It is interesting to note that the lifting of GIM cancellation in $D$-mixing by higher orders in the 
HQE \cite{Georgi:1992as,Ohl:1992sr,Bigi:2000wn,Bobrowski:2010xg} could also yield a sizeable 
$\CP$-violating phase in $\Gamma_{12}^D$, stemming from the second term on the r.h.s. of Eq.~(\ref{Gamma12D}).
According to Ref.~\cite{Bobrowski:2010xg} values of up to $1 \%$ for $\phi_{12}^D$ can currently not be excluded. Using 
Eq.~\ref{AGamma_Taylor} such a bound corresponds to $|A_\Gamma^{\rm SM}| \leq 2 \cdot 10^{-5}$.

An alternative origin of the discrepancy could lie in hypothetical  violations of quark-hadron duality. 
The success of the HQE in the bottom sector and in the lifetime of charmed mesons, where no GIM cancellations arise, seems to exclude huge duality-violation effects. On the other hand $D$-mixing might be more affected by duality violations than lifetimes due to the smaller number of hadronic states that contribute in oscillations.
In Ref.~\cite{Jubb:2016mvq} it was shown that duality violations as low as $20 \%$ could be 
sufficient to explain the discrepancy between experiment and the HQE prediction. 

Tt was pointed out very recently in Ref.~\cite{Lenz:2020efu} that the severe
GIM cancellations in the HQE prediction only arise if one uses identical renormalisation scales for box-diagram contributions related to internal $s \bar{s}$ 
($d \bar{d}$) and $s \bar{d}$ ($d \bar{s}$) quark pairs. 
Since the intermediate hadrons corresponding to $\Gamma^{D}_{ss,sd,dd}$ have either a net strangeness of zero ({\it e.g.} $K^+ K^-, \pi^+ \pi^-$) or one 
({\it e.g}. $K^+ \pi^-, \pi^+ K^-$) there seems to be no reason for having to choose the scales for these different states to be identical. 
Varying the renormalisation scales independently it is found that the HQE prediction also nicely encompasses the experimental result for $y$. To some extent this result  indicates that taking identical renormalisation scales for $\Gamma^{D}_{ss,sd,dd}$ in combination with the GIM cancellations implicitly assumes a precision of the individual contributions of the order of $10^{-5}$, which is clearly not realistic. Here a further study of higher order perturbative corrections  may provide additional insights (see Refs.~\cite{Asatrian:2017qaz,Asatrian:2020zxa} for first steps in this direction).

All in all there is  still more work to be done to obtain a deeper understanding of the HQE prediction for $\Gamma_{12}^D$, but it seems not unrealistic to hope that this approach may accommodate the measured values of the mixing parameters.
Once a satisfactory inclusive theory prediction for 
$\Gamma_{12}^D$  becomes available one could next aim for
a quark-level determination of $M_{12}^D$, or one could use dispersion relations as briefly discussed below. Having reliable predictions for both $\Gamma_{12}^D$
and $M_{12}^D$ we shall then be able to predict with confidence the weak mixing-phase $\phi_{12}^D$.

\subsubsection{Exclusive approach} \noindent
The exclusive approach, see {\it e.g.} Refs.~\cite{Wolfenstein:1985ft,Donoghue:1985hh,Falk:2001hx,Cheng:2010rv,Jiang:2017zwr}, 
aims to determine $M_{12}^D$ and  $\Gamma_{12}^D$ at the hadron level. A potential starting point is the expressions
\begin{eqnarray}
\Gamma_{12}^D & = & \sum \limits_n \rho_n \langle \Dzb | {\cal H}_{\rm eff}^{\Delta C =1} | n   \rangle
          \langle  n        | {\cal H}_{\rm eff}^{\Delta C =1} | \Dz \rangle \, ,
\\
M_{12}^D & = & \sum \limits_n \langle \Dzb | {\cal H}_{\rm eff}^{\Delta C =2} |\Dz \rangle 
+ {\cal P} \sum \limits_n \frac{ \langle \Dzb | {\cal H}_{eff.}^{\Delta C =1} | n   \rangle
\langle  n        | {\cal H}_{\rm eff}^{\Delta C =1} | \Dz \rangle }
                                          {m_D^2 - E_n^2} \, ,
\label{Dmix_exclusive}
\end{eqnarray}
where $n$ denotes all possible hadronic states into which both $\Dz$ and $\Dzb$ can decay, $\rho_n$ is the 
density of the state $n$ and ${\cal P}$ is the principal value. Unfortunately a first-principle calculation 
of the arising non-perturbative matrix elements is beyond our current abilities. Thus one has to make simplifying 
assumptions such as taking only the $SU(3)_F$-breaking effects due to phase space effects into account,
while neglecting any other hadronic effects, like $SU(3)_F$-breaking in the matrix elements.
Following this approach the authors of Ref.~\cite{Falk:2001hx} 
find that $y$ could naturally be of the
order of a per cent, in agreement with current measurements. In more detail, they
observe that contributions within a given
$SU(3)_F$-multiplet tend to cancel among
themselves and multi-particle final states
seem to give large contributions to mixing, while resonances close to the threshold are not very relevant.
Although encouraging, such a treatment 
clearly does not allow one to draw strong conclusions about the existence of BSM effects in the scenario that the 
measurement would disagree with these
expectations.  In Ref.~\cite{Cheng:2010rv,Jiang:2017zwr} this
method was improved by using experimental
input and by attempting to take into
account additional dynamical effects
beyond phase space. For example, in Ref.~\cite{Jiang:2017zwr} the
factorization-assisted topological-amplitude approach was used for this purpose, leading to the result
that two-particle final states can give a
value of $y = (0.21 \pm 0.07) \% $, which
might again indicate the importance of
multi-particle final states. 

The exclusive approach seems  naturally to 
predict a magnitude for $y$ in agreement with measurement. However, in order for this method to  be considered a genuine SM prediction further
improved experimental input (strong phases
and branching ratios), as well as a
theoretical inclusion of additional
$SU(3)_F$-breaking effects is necessary.
On a very long time-scale direct
lattice calculations might also become able to
predict the SM values for $D$-mixing, through
building on methods described in
Ref.~\cite{Hansen:2012tf}.
\subsubsection{Dispersion relation} 

A dispersive relation between $x$ and $y$ was derived in Ref.~\cite{Falk:2004wg} in the heavy quark limit:
\begin{equation}
    \Delta M_D = - \frac{1}{2 \pi} {\cal P}
    \int \limits_{2 m_\pi}^{\infty}
    dE \left[ \frac{\Delta \Gamma_D(E)}{E-M_D} + {\cal O} \left( \frac{\Lambda_{QCD}}{E}\right)
    \right]\, .
\end{equation}
The authors conclude that if $y$ is in the
ballpark of $+1\%$, then one expects $x$ to be in the range of $10^{-3}$ to $10^{-2}$, again in agreement with current measurements.
Very recently this dispersion relation, in combination with additional assumptions, has been
used to derive bounds on $\CP$ violation in $D$-mixing \cite{Li:2020xrz}.

\subsection{Experiment}

The existence of charm mixing was first established in 2007 from combining the ensemble of results from the $B$-factories, CDF and earlier experiments~\cite{Amhis:2019ckw}.  The first single-experiment observation was made by LHCb in 2013~\cite{Aaij:2012nva}, and was followed by CDF and Belle soon after~\cite{Aaltonen:2013pja,Ko:2014qvu}.

Mixing studies may be divided into three main categories: measurements of decays into $\CP$ eigenstates, measurements of decays into  flavour eigenstates, and analyses of multi-body self-conjugate modes that take account of the position of the decay in phase space.  

\subsubsection{Measurements with $\CP$ eigenstates}

Charm mixing in $\Dz$ decays to $\CP$ eigenstates of eigenvalue $\eta_{\CP}$ gives rise to an effective lifetime $\tau_{\CP}$, which differs from that in decays to flavour-specific states, $\tau_{\rm FS}$.  The observable 
\begin{equation}
    y_{\CP} \equiv \eta_{\CP}\left(\frac{\tau_{\rm FS}}{\tau_{\CP}} -1 \right),
\label{eq:ycp_def}
\end{equation}
is equal to $y$ in the limit of $\CP$ conservation and more generally is given by Eq.~\ref{eq:ydef1}. When allowing for direct $\CP$ violation, specific to mode $f$ with decay amplitude $A_f$, then Eq.~\ref{eq:ydef1} receives an additional contribution of $(A_D^f/2) x\sin\phi^D$, with $A_D^f \equiv |A_f/\bar{A}_{{f}}|^2 - 1$. 

Valuable information may also be gained from  
studying the time-dependent asymmetry
\begin{equation}
    {{A}}_{\CP}\left(f;t\right) \equiv \frac{\Gamma\left(\Dz(t)\to f\right) - \Gamma\left(\Dzb(t)\to f\right)}{\Gamma\left(\Dz(t)\to f\right) + \Gamma\left(\Dzb(t)\to f\right)},
    \label{eq:ACP}
\end{equation}
which in the case of $\CP$ violation has an approximate linear dependence 
\begin{equation}
   {{A}}_{\CP}\left(f;t\right) \approx  a^{\rm dir}_{\CP}(f) - A_\Gamma(f) \frac{t}{\tau}
\end{equation}
driven by a potential direct $\CP$-violating asymmetry $a^{\rm dir}_{\CP}(f)$, and a second parameter $A_\Gamma(f)$. When direct $\CP$ violation is neglected the latter contribution is universal, and is given by Eq.~\ref{eq:agdef1}, otherwise this expression receives an offset of $-y a^{\rm dir}_{\CP}(f)$.

\begin{figure}[htb]
    \centering
    \includegraphics[width=0.85\textwidth]{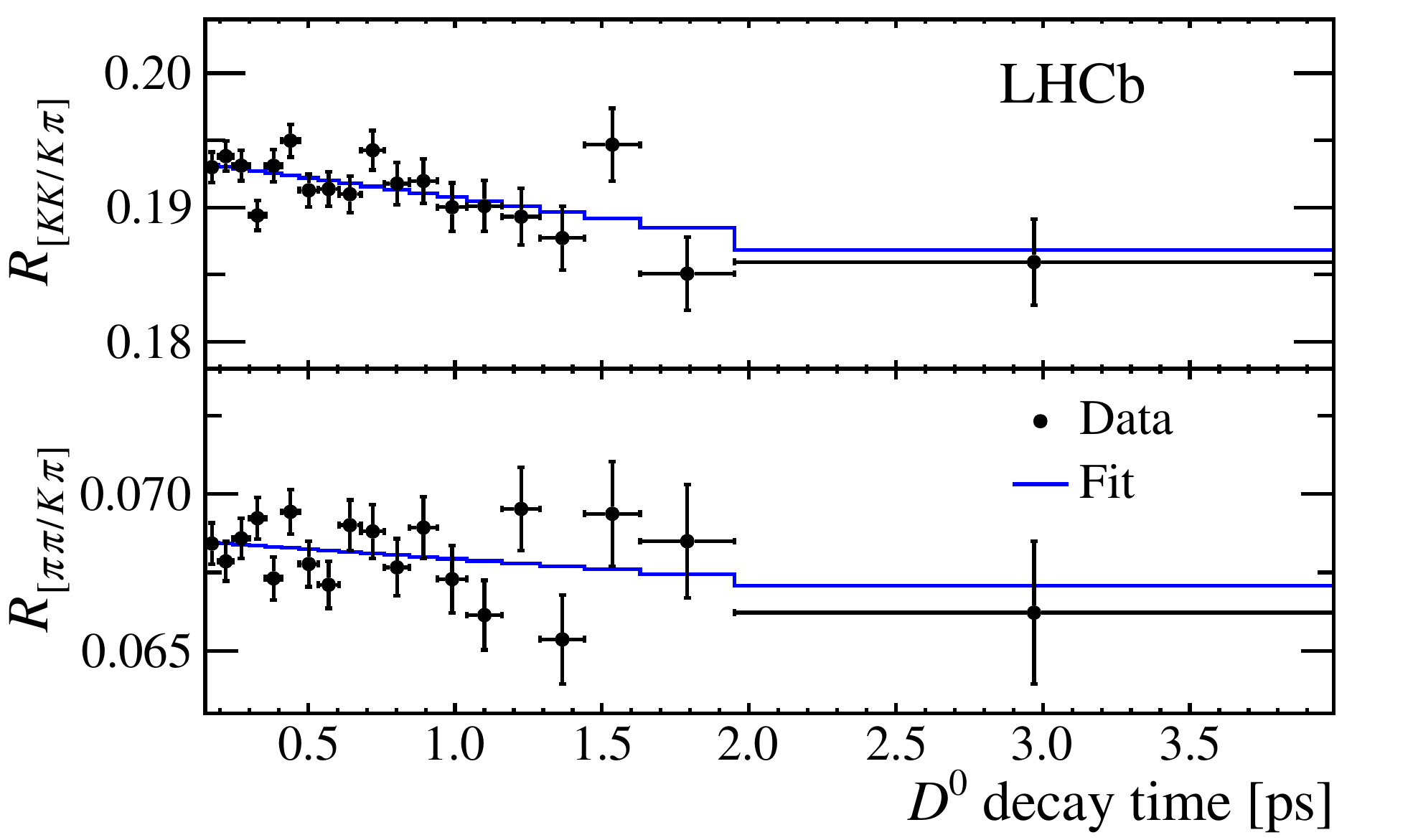}
    \caption{Ratio of the lifetime distributions of $\Dz\to K^+K^-$ to $\Dz \to K^-\pi^+$ and $\Dz\to \pi^+\pi^-$ to $\Dz \to K^-\pi^+$ decays~\cite{Aaij:2018qiw}.}
    \label{fig:ycp}
\end{figure}

Measurements  of $y_{\CP}$ and $A_\Gamma$ have been performed by several experiments, usually exploiting the decays $\Dz \to K^+K^-$ and  $\Dz \to \pi^+\pi^-$.  A selection of the most precise results is presented in Table~\ref{tab:ycpAG}. With current sensitivity there is no indication of a difference between $A_\Gamma(K^+K^-)$ and $A_\Gamma(\pi^+\pi^-)$, and so the results are combined to yield a universal observable.  
Figure~\ref{fig:ycp} shows the ratio of the lifetime distributions of $\Dz \to K^+K^-$ to $\Dz \to K^-\pi^+$, and $\Dz \to \pi^+\pi^-$ to $\Dz \to K^-\pi^-$ from Ref.~\cite{Aaij:2018qiw} in which the non-zero slope is clearly evident.

\begin{table}[htb]
        \caption{Selected results for $y_{\rm CP}$ and $A_\Gamma$. All values  derive from the combination of $\Dz \to K^+K^-$ and $\pi^+\pi^-$, apart from Ref.\,\cite{Zupanc:2009sy} which is from $\Dz \to K^0_{\rm S}K^+K^-$. The combination for Ref.\,\cite{Aaij:2019yas} has been made by the authors of this review. When two uncertainties are shown, they are statistical and systematic, respectively.}
    \centering
    \begin{tabular}{lclc}\hline
    Measurement & \multicolumn{1}{c}{$y_{\CP}$ (\%)} &  Measurement & \multicolumn{1}{c}{$A_{\Gamma}$ (\%)} \\ \hline
        Belle, 673\,fb$^{-1}$\,\cite{Zupanc:2009sy} & $0.11 \pm 0.61 \pm 0.52$ &  Belle, 976\,fb$^{-1}$\,\cite{Staric:2015sta}  & $-0.03 \pm 0.20 \pm 0.07$ \\
    Belle, 976\,fb$^{-1}$\,\cite{Staric:2015sta} & $1.11 \pm 0.22 \pm 0.09$ &  CDF, 9.7\,fb$^{-1}$\,\cite{Aaltonen:2014efa} & $-0.12 \pm 0.12$ \\
    BaBar, 468\,fb$^{-1}$\cite{Lees:2012qh} & $0.72 \pm 0.18 \pm 0.12$ &  LHCb $D^{\ast +}$, 3\,fb$^{-1}$\,\cite{Aaij:2017idz} & $-0.013 \pm 0.028 \pm 0.010$ \\
    LHCb  SL, 3\,fb$^{-1}$\,\cite{Aaij:2018qiw} & $0.57 \pm 0.13 \pm 0.09$ &  LHCb SL, 5.4\,fb$^{-1}$\,\cite{Aaij:2019yas} & $-0.029 \pm 0.032 \pm 0.004$  \\   
   World average\,\cite{Amhis:2019ckw} & $0.72 \pm 0.11$ &  World average\,\cite{Amhis:2019ckw} & $-0.031 \pm 0.020$ \\   \hline
    \end{tabular}
    \label{tab:ycpAG}
\end{table}

The measurement of $y_{\CP}$ in a hadron collider is challenging. The trigger generally introduces a decay-time dependent acceptance that is not identical for $\Dz \to K^+K^-$, $\pi^+\pi^-$ and $K^-\pi^+$, on account of the different masses of the decay products.  The most precise measurement to emerge so far from LHCb is for semileptonically tagged decays with Run 1 data, where this problem is less acute, but for which the statistical and systematic uncertainties are still of a similar magnitude.  Currently $y_{\CP}$ is measured with a precision of 0.11\%, and is compatible with the measurements of $y$ from other methods.

The determination of $A_{\Gamma}$ is inherently robust, as an asymmetry measurement involves the same final state, and the parameter of interest is the slope of the asymmetry, which can only be biased by time-dependent systematic effects in the tagging process, residual backgrounds or secondary contamination.  The experimental picture is 
now wholly dominated by LHCb measurements, which themselves are limited by the statistical uncertainties.  The measurement procedure is validated by determining a `pseudo $A_{\Gamma}$' with the Cabibbo-favoured $\Dz \to D^-\pi^+$ decay, where negligible \CP\ violation, and hence a result of zero, is expected.  Self-consistency is also required over all running periods, in particular between those of opposite dipole polarity.  The current world average for $A_{\Gamma}$ is compatible with $\CP$ conservation, and is measured with a precision of $2 \times 10^{-4}$, which makes it one of the best known quantities in flavour physics.  Significant near-term improvements can be expected from the analysis of the LHCb Run 2 prompt sample.

Other decay modes can be harnessed for these measurements.  Belle has studied the $\CP$-odd channel $\Dz \to K^0_{\rm S}\omega$~\cite{Nayak:2019byo}, and also  performed a measurement that compares the time evolution of different regions of the Dalitz plot for the decay $\Dz \to K^0_{\rm S}K^+K^-$~\cite{Zupanc:2009sy}.   It has been proposed that self-conjugate channels such as $\Dz \to \pi^+\pi^-\pi^0$ can also be exploited in an inclusive manner, provided their net $\CP$ content is known~\cite{Malde:2015xra}.  Modes such as $\Dz \to K^0_{\rm S}\pi^0$ are unsuitable, even at $B$-factories, because of the poor resolution on the proper time. 

\subsubsection{Measurements with flavour eigenstates}
\label{sec:flaveigen}

Decays into flavour-specific final states, in particular those that can be accessed by both CF and DCS amplitudes,
play a central role in charm-mixing analyses.
The most studied of these channels is the so-called {\it wrong-sign} decay $\Dz \to K^+\pi^-$, to be distinguished from the {\it right-sign} decay $\Dz \to K^-\pi^+$.

In the absence of $\CP$ violation the time-dependent ratio between wrong-sign and right-sign decays is
\begin{equation}
    R(t/\tau) \approx R_D + \sqrt{R_D}y'(t/\tau) + \frac{x^{'2} + y^{'2}}{4}({t/\tau})^2\,.
\label{eq:wskpi}
\end{equation}
Here the first term, $R_D \approx 0.34\%$, is the squared ratio of the DCS to CF amplitudes, the second term arises from interference between the mixing and the DCS amplitudes,
and the final term is generated by the mixing amplitude alone. The parameters
$x{'} = x\cos\delta + y\sin\delta$ and $y{'} = y\cos\delta - x\sin\delta$, where $\delta$ is the strong-phase difference between the CF and DCS amplitudes, the value of which must be known in order to interpret the results in terms of the mixing parameters.  Although this strong-phase difference may be measured in quantum-correlated decays at charm threshold, the uncertainty on our current knowledge, $\delta=(16.1^{+7.9}_{-10.1})^\circ$, is essentially set by a global fit to all charm-mixing measurements~\cite{Amhis:2019ckw}.  Future analyses exploiting larger $\psi(3770)$ data sets, and over-constrained fits to $b$-physics measurements made for the purpose of determining the Unitarity Angle $\gamma$, which also exploit this charm decay, will help to improve the precision on the strong-phase difference.

The analysis may be extended to probe for $\CP$ violation by measuring $R(t/\tau)^+$ and $R(t/\tau)^-$, the separate ratios for $\Dz$ and $\Dzb$ decays, respectively.  Direct $\CP$ violation, here expected to be negligible in the Standard Model, would generate different values of the offset, and indirect $\CP$ violation would change the time-dependent terms, with $x'^\pm = |q/p|^{\pm 1}(x'\cos\phi^D\pm y' \sin \phi^D)$ and $y'^\pm = |q/p|^{\pm 1}(y'\cos\phi^D \mp x' \sin \phi^D)$.

\begin{figure}[htb]
    \centering
        \includegraphics[width=0.55\textwidth]{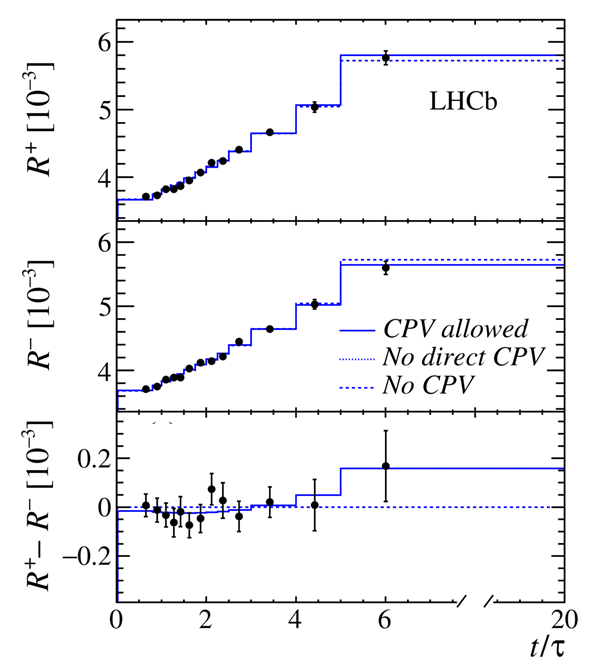}
    \caption{The ratio $R(t/\tau)$ in the $K\pi$ mixing analysis, shown for $\Dz$- and $\Dzb$-tagged decays ($R^+$ and $R^-$, respectively), and the difference between them~\cite{Aaij:2017urz}. Various fit hypotheses are superimposed.}
    \label{fig:kpiWS}
\end{figure}

Results from selected measurements of the effective mixing parameters are shown in Table~\ref{tab:kpiws}.  The most precise study, from which the $R(t/\tau)$ distributions are shown in Fig.~\ref{fig:kpiWS}, is still limited by the statistical uncertainty~\cite{Aaij:2017urz}.  The dominant systematic uncertainties are determined from the data themselves, and so are expected to decrease with future measurements.  For example, the wrong-sign sample suffers contamination from right-sign decays in which the kaon is misidentified as a pion, and the pion as a kaon.  Knowledge of the magnitude of this background is limited by the understanding of the particle-identification performance, which is studied using control samples in data.  The parameter $y'$ is currently measured with a precision of about $0.05\%$, and $x'^{2}$ is compatible with zero.  Results of separate fits to $R^+(t/\tau)$ and $R^-(t/\tau)$ 
are consistent with $\CP$ conservation. 

\begin{table}[htb]
\caption{Results, assuming $\CP$ conservation, of the effective mixing parameters from selected measurements of the time-dependent $\Dz \to K^+\pi^-$ to $\Dz \to K^-\pi^+$ ratio performed at the $B$-factories, Tevatron and LHC. When two uncertainties are shown, they are statistical and systematic, respectively.}
    \centering
    \begin{tabular}{lcc} \hline
    & $y'$ (\%) & $x'^{2}$ ($10^{-3}$) \\ \hline
    Belle, 976\,fb$^{-1}$\,\cite{Ko:2014qvu} & $0.46 \pm 0.34$ & $0.09 \pm 0.22$ \\
    CDF, 9.6\,fb$^{-1}$\,\cite{Aaltonen:2013pja}& $0.43 \pm 0.43$ & $0.08 \pm 0.18$ \\
    LHCb $D^{*+}$, 5.0\,fb$^{-1}$\,\cite{Aaij:2017urz} & $0.528 \pm 0.045 \pm 0.027$ & $0.039 \pm 0.023 \pm 0.014$ \\ \hline
    \end{tabular}
    \label{tab:kpiws}
\end{table}

It is possible to analyse multi-body CF/DCS decays, such as $\Dz \to K^\mp\pi^\pm\pi^+\pi^-$ and $\Dz \to K^\mp\pi^\pm\pi^0$ in a similar fashion. In this case, however, the interpretation of the time-dependent wrong-sign to right-sign ratio must take account of the fact that the amplitude ratio and strong-phase difference varies over the phase space of the final-state particles. In an inclusive analysis the equivalent parameters for $R_D$ and $\delta$ are averaged over this phase space for the decay in question~\cite{Harnew:2013wea}.  Furthermore, the linear term in $y'$ of Eq.~\ref{eq:wskpi} is multiplied by a parameter called the {\it coherence factor}~\cite{Atwood:2003mj}.  The coherence factor takes a value between zero and unity and dilutes the mixing signature.  Both the coherence factor and the average strong-phase difference can be determined at charm threshold~\cite{Lowery:2009id,Evans:2016tlp} and used to interpret the multibody mixing measurement in terms of $x$ and $y$, as has been demonstrated in an LHCb study of $\Dz \to K^{\mp}\pi^{\pm}\pi^+\pi^-$ decays~\cite{Aaij:2016rhq}.  The dilution and consequent loss of information that comes with an inclusive analysis can be partially recovered through performing the measurement in bins of phase space~\cite{Evans:2019wza}.
Full statistical sensitivity can be achieved by using models to track the continuous variation of the phase difference between CF and DCS amplitudes in an unbinned measurement, as was shown by the BaBar collaboration in a study of $\Dz \to K^\mp\pi^\pm\pi^0$ decays~\cite{Aubert:2008zh}.   However, the systematic uncertainty arising from imperfections in the models is very difficult to assess, rendering this approach unsatisfactory for precision studies.

\subsubsection{Measurements with multi-body self-conjugate decays}

\begin{figure}
    \centering
    \includegraphics[width=0.99\textwidth]{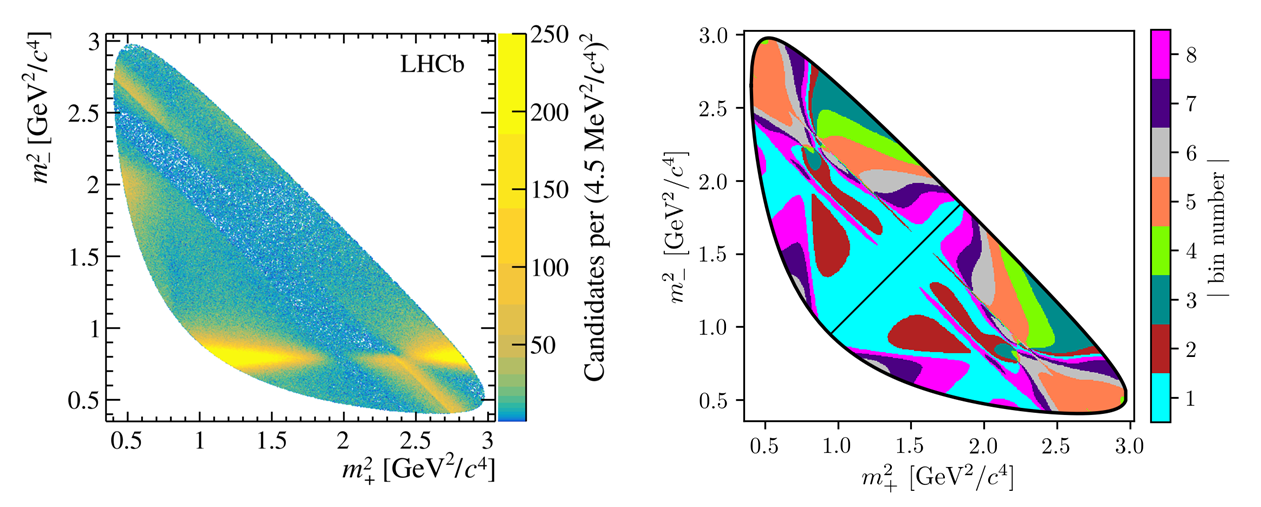} 
    \caption{Dalitz plot of $\Dz \to K^0_{\rm S}\pi^+\pi^-$ decays, with $m_\pm^2 = m^2(K^0_{\rm S}\pi^\pm)$.  Left: data~\cite{Aaij:2019jot}). Right: bins of equal-interval strong-phase difference as defined in Ref.~\cite{Libby_2010}.} 
    \label{fig:kspipi}
\end{figure}

As is evident from its Dalitz plot, shown in Fig.~\ref{fig:kspipi}~(left), the mode $\Dz \to K^0_{\rm S}\pi^+\pi^-$ has a rich resonance structure. The decay receives contributions both from $\CP$ eigenstates, such as $\Dz \to K_{\rm S}^{0} \rho^0$, and flavour-eigenstates such as $\Dz \to K^{\ast\mp}\pi^\pm$.  Hence, as first pointed out in Ref.~\cite{Asner:2005sz}, an analysis that pays attention to the position of each decay in phase space can be considered as a combination of the two strategies discussed above. Moreover, the strong-phase difference varies over the Dalitz plot, ensuring that the analysis has greater sensitivity to the parameter $x$ than in the case of $\Dz \to K^\mp\pi^\pm$ decays. The measurement may be performed separately for $\Dz$ and $\Dzb$ mesons to probe for $\CP$ violation.  There is sufficient information in the Dalitz plot to allow a simultaneous fit of $x$, $y$, $\phi^D$ and $|q/p|$, giving $\Dz \to K^0_{\rm S}\pi^+\pi^-$ a particularly important role in charm studies.  

Knowledge of the strong-phase variation is available from two alternative sources, following similar considerations to those discussed for multi-body decays in Sect.~\ref{sec:flaveigen}.
Firstly, an amplitude model may be constructed from a fit to the time-integrated Dalitz plot to describe the contribution of each resonance and the interference between them. A second approach is to use charm-threshold data to measure the mean strong-phase difference (or, in practice, the amplitude-weighted cosine and sine of this quantity) in bins of phase space.  The definition of these bins is itself guided by amplitude models. Figure~\ref{fig:kspipi}~(right) shows one such partitioning of the Dalitz plot, in which eight pairs of   symmetrical bins are chosen. The bin boundaries separate the expected phase difference into intervals of $2\pi/8$ radians.  Measurements of the strong-phase difference within these bins have been performed by both CLEO and BESIII~\cite{Libby_2010,Ablikim:2020lpk}.

The mixing analysis is performed through either an unbinned fit of the data set that assumes the amplitude variation given by the model, or a binned analysis that makes use of the charm-threshold inputs~\cite{Bondar:2010qs,Thomas:2012qf}.   The latter strategy has the benefit of incurring no model dependence, at the expense of some loss of statistical sensitivity.  At hadron-collider experiments the trigger can induce significant acceptance variations over the Dalitz plot, which may also have a dependence on proper time.  To ameliorate this problem a modified version of the model-independent binned analysis has been proposed, in which ratios of yields are constructed in pairs of symmetric bins, much like the $R(t/\tau)$ ratio of Eq.~\ref{eq:wskpi}.  This {\it bin-flip method} is experimentally robust, and also has the consequence of enhancing sensitivity to the less well-known mixing parameter $x$, albeit at the expense of degrading sensitivity to $y$~\cite{DiCanto:2018tsd}.

Results are shown in Table~\ref{tab:kshh} for model-dependent analyses from the $B$-factories, and a model-independent bin-flip analysis from LHCb.  
The sample size exploited in the LHCb measurement is around 2.3 million decays, approximately four times that of each of the $B$-factory studies. This analysis provides the single most precise determination of the parameter $x$. The systematic uncertainties on the mixing parameters are around a quarter of the statistical uncertainties, and are dominated by the finite knowledge of the strong-phase information, which in this study comes from Ref.~\cite{Libby_2010} alone.
Other self-conjugate multi-body decays that have been exploited for charm-mixing measurements include $\Dz \to K^0_{\rm S}K^+K^-$~\cite{delAmoSanchez:2010xz} and $\Dz \to \pi^+\pi^-\pi^0$~\cite{TheBABAR:2016gom}.

\begin{table}[]
\caption{Results for the mixing and the $\CP$-violation parameters from $\Dz \to K^0_{\rm S}\pi^+\pi^-$ decays.  The results for Ref.~\cite{delAmoSanchez:2010xz} also includes input from $\Dz \to K^0_{\rm S}K^+K^-$.}
    \centering
    \begin{tabular}{lccc} \hline
    & Belle,  &  BaBar, & LHCb $D^{\ast +}$ and  \\
    & 540\,fb$^{-1}$\,\cite{Abe:2007rd} & 469\,fb$^{-1}$\,\cite{delAmoSanchez:2010xz} &  SL, 3.0\,fb$^{-1}$\,\cite{Aaij:2019jot} \\ \hline
    $x$ (\%)     & $0.81^{+0.33}_{-0.35}$ & $0.16 \pm 0.27$& $0.27^{+0.17}_{-0.15}$\\ 
    $y$ (\%)     & $0.37^{+0.27}_{-0.29}$ & $0.57\pm 0.25$ & $0.74 \pm 0.37$\\
    $\phi^D$ $(^\circ)$     &  $-14^{+17}_{-18}$ & / & $-5.2^{+6.3}_{-9.2}$\\
    $|q/p|$ & $0.86^{+0.32}_{-0.30}$ & / & $1.05^{+0.22}_{-0.17}$\\ \hline
    \end{tabular} 
    \label{tab:kshh}
    \end{table}
    
\subsubsection{Summary of current knowledge and experimental prospects}

\begin{figure}
    \centering
    \includegraphics[width=0.9\textwidth]{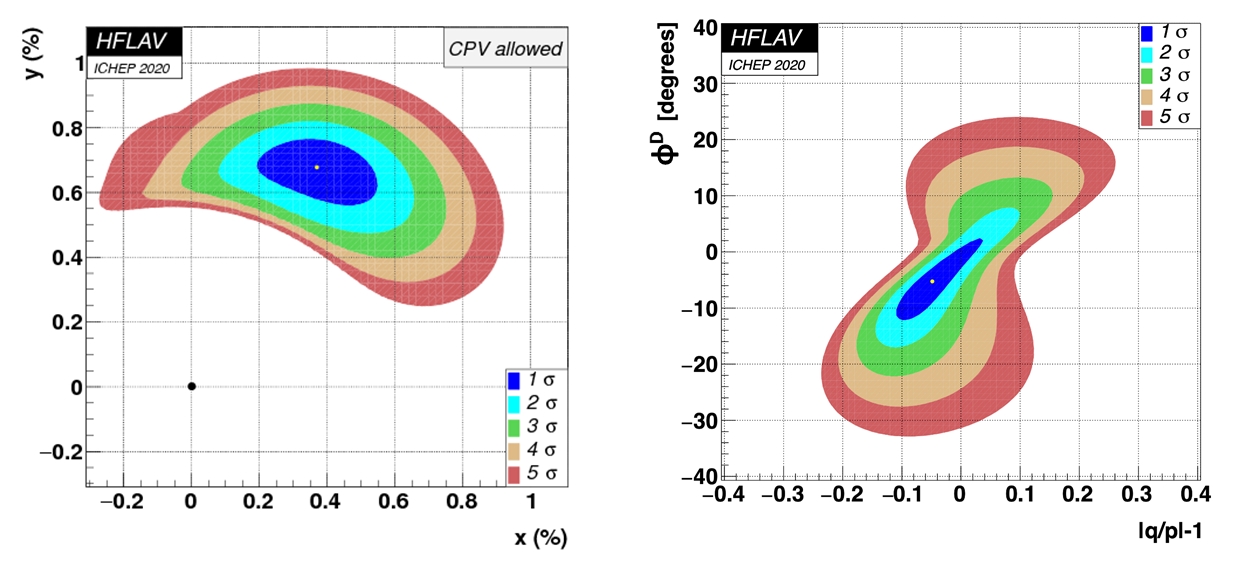}
    \caption{Contours for the mixing parameters $x$ and $y$, and for the $\CP$-violation parameters $\phi^D$ and $|q/p|-1$~\cite{Amhis:2019ckw}.}
    \label{fig:hflav}
\end{figure}

Fits are performed to the ensemble of charm-mixing data by the HLAV group, with the most recent update coming in July 2020~\cite{Amhis:2019ckw}.  The central values and one-sigma uncertainties on the mixing parameters are
\begin{eqnarray}
x  = (0.37\pm 0.12) \% \, ,\:\:
&& y  = 0.68^{+0.06}_{-0.07} \%  \, ,\nonumber \\
|q/p|=0.951^{+0.053}_{-0.042}\, , \:\: && \phi^D=(-5.3^{+4.9}_{-4.5})^\circ\, ,
\label{eq:xyvalues}
\end{eqnarray}
and the one- to five-sigma contours are displayed in Fig.~\ref{fig:hflav}.  In summary, $y$ is well measured, but a non-zero value of $x$ is not yet excluded.  
The measurements do not yet have sufficient sensitivity to approach the Standard Model expectation for $CP$ violation, but already confirm that any effects are small.    

Steady progress can be expected over the coming two decades.  A complete set of analyses based on the full Run 1 and 2 LHCb data sets will already bring a significant increase in sensitivity. The samples will grow by an order of magnitude at Upgrade~I of LHCb, and important measurements will soon arrive from  Belle~II.  At LHCb Upgrade~II the statistical uncertainties on $|q/p|$ and $\phi^D$ are expected to shrink to 0.004 and 0.18$^\circ$, respectively, from the $\Dz \to K^0_{\rm S}\pi^+\pi^-$ analysis alone~\cite{lhcbupgrade2}.  The challenge will be 
to attain sufficient experimental control to match these advances in statistical precision.  It will also be necessary to improve the measurements of strong-phase parameters, for example at a Super Tau Charm Factory, although it will become possible to also fit these parameters with reasonable sensitivity as part of the next generation of LHCb and Belle~II analyses.

 \subsection{Bounds on BSM models from $D$ mixing}
Because of its pronounced SM
suppression, $D$-mixing is
particularly suited for indirect searches for BSM effects. This attribute is exploited in Ref.~\cite{Golowich:2007ka}, where a wide
class of new physics models is
discussed.
Assuming an off-diagonal tree-level coupling of some BSM mediator to the neutral $D$ meson, one is able  test BSM scales higher than $10^4$\,TeV (see {\it e.g.} Ref.~\cite{Silvestrini:2015kqa} based on Ref.~\cite{Bona:2007vi}).
It is common in such BSM studies to neglect the SM contribution to $D$-mixing and to saturate the experimental results completely by the BSM contribution.
Without doubt, $D$-mixing measurements  give very stringent constraints on any BSM model that acts in the charm sector, as is also stressed below in Sect.~\ref{subsub:BSM} in the context of direct $\CP$ violation.


 \section{Direct \CP\ violation in singly Cabibbo-suppressed decays}
\label{sec:direct}

The most promising avenue to search for direct $\CP$ violation in charm is through the study of SCS decays, on account of the contribution of penguin diagrams to these processes. 
Although not discussed further here, it should be stressed that CF and DCS decays are also of interest, as any signal of direct $\CP$ violation here would be a clear sign of BSM effects.
Another worthwhile area of investigation is the study of $\CP$-violating effects in flavour-changing neutral-current charm decays, as discussed in Refs.~\cite{Bause:2020obd,Fajfer:2012nr}, but this topic is beyond the scope of this review.

In this section we briefly review the status of two-body SCS $\CP$-asymmetry measurements and the experimental challenges that these entail, before focusing in more detail on the determination and interpretation of $\Delta A_{\CP}$, the difference in the time-integrated $\CP$ asymmetries of $\Dz \to K^+K^-$ and $\Dz \to \pi^+\pi^-$.  We then give a summary of the methods and status of $\CP$ studies of multi-body SCS decays.  We conclude with some remarks on $\CP$-violation measurements in the charm-baryon sector. 

\subsection{Measuring the $\CP$ asymmetry and studies of two-body decays}

For charged mesons and baryons the $\CP$ asymmetry $A_{\CP}(f)$ is the time-independent form of Eq.~\ref{eq:ACP}, with $f$ now not restricted to $\CP$ eigenstates, and $A_{\CP}(f) = a_{\rm dir}(f)$, which is the direct $\CP$ asymmetry for the mode in question.  For neutral mesons $A_{\CP}(f)$ is integrated with respect to proper time and, in addition to $a_{\rm dir}$, receives a contribution from indirect $\CP$ violation.

In general the measured asymmetry $A_{\rm raw}(f)$ does not exactly correspond  to the underlying physics asymmetry, as it can be biased by several detector- and facility-related effects.  In the case that the $\CP$ asymmetry and these effects are small, then
\begin{equation}
    A_{\rm raw}(f) = A_{\CP}(f) + a_{\rm det}(f) + a_{\rm tag}(f) + a_{\rm prod}(f) .
\end{equation}
Here $a_{\rm det}(f)$ is the detection asymmetry for final state $f$, which can be non-zero for any decay apart from those into a $\CP$ eigenstate involving pseudoscalars.  Contributions to   $a_{\rm det}(f)$ come from, for example, differences in trigger or reconstruction efficiency for positive and negative particles, or from the different interaction cross-sections in the detector material for kaons with positive and negative strangeness.  The tagging asymmetry $a_{\rm tag}(f)$ applies only for measurements with neutral mesons and is driven by different detection efficiencies for the slow pion or lepton tags. Although these effects are independent of the final state $f$, correlations can occur through the acceptance of the detector.  Finally, the $pp$ initial state at the LHC can lead to a  production asymmetry $a_{\rm prod}(f)$, with different proportions of hadrons of each flavour being produced within the detector acceptance.  Again, correlations induced by detector acceptance effects can result in the effective value of this asymmetry acquiring a dependence on the final state. Typically $a_{\rm det}(f)$, $a_{\rm tag}$ and $a_{\rm prod}(f)$ are all ${\cal{O}}(\%)$, which is larger than the expected value of $A_{\CP}(f)$.

When measuring $A_{\CP}$ for a SCS decay it is customary to use CF control channels to account for the detection, tagging and production asymmetries.  Negligible $CP$ violation is expected for these modes, meaning that any non-zero raw asymmetry must arise from the other sources.  For example, in the $D^{\ast +}$-tagged determination of $A_{\CP}(K^+K^-)$ at LHCb measurements of the raw asymmetries of the modes $\Dz \to K^-\pi^+$, $D^+ \to K^-\pi^+\pi^+$ and $D^+ \to \bar{K^0}\pi^+$ are employed to subtract the detection and production asymmetries~\cite{Aaij_2017}. These asymmetries are measured after re-weighting the kinematics of the control samples so that they match those of the quantities they are being used to describe. Furthermore, fiducial cuts are applied to exclude regions of the acceptance where detector asymmetries are known to be large.  Finally, a small correction must be made for cross-section interaction effects, mixing, and $\CP$ violation for the neutral kaon in the  third control mode, but this may be calculated with excellent precision. 

Table~\ref{tab:acpscs} lists the most precise individual measurements of the $\CP$ asymmetries for SCS decays of $\Dz$, $D^+$ and $D_s^+$ mesons into two-body final states involving pions and kaons (note that a more precise determination of $A_{\CP}(\pi^+\pi^-)$ can be obtained through combination of $A_{\CP}(K^+K^-)$ and the $\Delta A_{\CP}$ measurement, discussed below). All results are compatible with $\CP$ conservation.  The precision of these measurements is expected to improve with the analysis of larger samples, as many of the contributions to the systematic uncertainties are directly determined from the data themselves.

\begin{table}[]
\caption{Most precise individual measurements of $A_{\CP}$ in two-body SCS decays. 
The measurements from Ref.~\cite{Aaij:2019vnt} are corrected for the effects of $\CP$ violation in the neutral kaon system. The uncertainties are statistical and systematic, respectively.}
    \centering
    \begin{tabular}{lcc} \hline
    Mode & $A_{\CP}$ (\%) & Experiment \\ \hline
    $\Dz \to \pi^+\pi^-$ & $+0.22 \pm 0.24 \pm 0.11$ & CDF, 5.9\,fb$^{-1}$\,\cite{Aaltonen:2011se}\\
    $\Dz \to K^+K^-$ & $+0.14 \pm 0.15 \pm 0.10$  &  LHCb $D^{\ast +}$ and SL, 3.0\,fb$^{-1}$\,\cite{Aaij_2017}\\
    $\Dz \to \pi^0\pi^0$ & $-0.03 \pm 0.64 \pm 0.10$ & Belle, 966\,fb$^{-1}$\,\cite{Nisar:2014fkc} \\
    $\Dz \to K^0_{\rm S}K^0_{\rm S}$ & $-0.02 \pm 1.53 \pm 0.17$ &Belle, 921\,fb$^{-1}$\,\cite{Dash:2017heu} \\
    $D^+ \to \pi^+\pi^0$ & $+2.31 \pm 1.24 \pm 0.23$ &  Belle, 921\,fb$^{-1}$\,\cite{Babu:2017bjn}  \\
    $D^+ \to K^0_{\rm S}K^+$ & $-0.009 \pm 0.065 \pm 0.048$ & LHCb, 3.8\,fb$^{-1}$\,\cite{Aaij:2019vnt} \\
    $D_s^+ \to K^+ \pi^0$ & $-26.6 \pm 23.8 \pm 0.9$ & CLEO, 586\,nb$^{-1}$\,\cite{Mendez:2009aa} \\
    $D_s^+ \to K^0_{\rm S}\pi^+$ & $+0.13 \pm 0.19 \pm 0.05$ & LHCb, 3.8\,fb$^{-1}$\,\cite{Aaij:2019vnt} \\ \hline
    \end{tabular}
    \label{tab:acpscs}
\end{table}

\subsection{The measurement of $\Delta A_{CP}$}

\subsubsection{The measurement and future prospects} 

The quantity $\Delta A_{\CP} \equiv A_{\CP}(K^+K^-) - A_{\CP}(\pi^+\pi^-)$ can be measured with very small systematic uncertainty. If the variation in acceptance of both decays is the same, an assumption that can be made valid by re-weighting the kinematics of the two samples to agree, then the tagging and production asymmetries of the decays are also identical, and so  $\Delta A_{\CP} = A_{\rm raw}(K^+K^-) - A_{\rm raw}(\pi^+\pi^-)$. 

LHCb has performed this measurement  with 5.9\,fb$^{-1}$ of data collected in Run~2 using both $D^{\ast +}$ and semileptonic tags~\cite{Aaij:2019kcg}, and obtained
\begin{eqnarray}
    \Delta A_{\CP}\,(D^{\ast +}) &=& \left[ -18.2 \pm 3.2 \,{\rm (stat.)} \pm 0.9\, {\rm (syst.)}\right] \times 10^{-4} \nonumber \\
    \Delta A_{\CP} \,(SL) &=& \left[ -9 \pm 8 \,{\rm (stat.)} \pm 5\, {\rm (syst.)} \right] \times 10^{-4}, \nonumber 
\end{eqnarray}
which can be combined with measurements made with 3.0\,fb$^{-1}$ of data from Run~1~\cite{Aaij:2014gsa,Aaij:2016cfh} to give
\begin{eqnarray}
    \Delta A_{\CP} &=& (-15.4 \pm 2.9)\times 10^{-4}. \nonumber
\end{eqnarray}
The significance of the deviation from zero is 5.3 standard deviations, and hence this measurement constitutes the first observation of $\CP$ violation in charm decays.  It is instructive to note the enormous samples required for this important result: around 51 million (11 million) $D^{\ast +}$ (SL) -tagged  $\Dz \to K^+K^-$ decays and 11 million (4 million)  $\Dz \to \pi^+\pi^-$ decays.

Because of the presence of mixing, there is the possibility of a contribution from indirect $\CP$ violation to the individual asymmetries. Therefore
\begin{equation}
    \Delta A_{\CP} = \Delta a_{\rm dir} - \frac{\Delta \langle t \rangle}{\tau}A_\Gamma,
\end{equation}
where $\Delta a_{\rm dir}$ is the difference in direct $\CP$ asymmetries between the two decays, $\Delta\langle t \rangle $ is the difference in the mean decay times of the two decays in the analysis, and $\tau$ is the $\Dz$ lifetime.  This expression assumes that the indirect $\CP$ violation, manifested through $A_\Gamma$, is the same for the two decays.  It is then found that
\begin{equation}
 \Delta a_{\rm dir} = (-15.7 \pm 2.9) \times 10^{-4} , 
 \label{ACP_dir_Exp}   
\end{equation}
indicating that $\Delta A_{\CP}$ is primarily sensitive to direct $\CP$ violation, and that the direct $\CP$ asymmetries are different for the two decays.

There are no known irreducible systematic uncertainties in the $\Delta A_{\CP}$ measurement, so significant improvements in knowledge can be expected at LHCb over the coming years, with the statistical precision after Upgrade~II predicted to be an order of magnitude better than at present~\cite{lhcbupgrade2}.  It is naturally of great interest to know the values of the individual asymmetries, and an improved measurement of $A_{\CP}(K^+K^-)$ is eagerly awaited from the full Run~2 data set. Most likely, however, data from Run~3 will be required to establish a non-zero asymmetry for either decay.
The final sample sizes at Belle~II for the two modes will be smaller than those already accumulated by LHCb, however measurements performed there will be valuable on account of  the very different experimental environment.

\subsection{Theory for $\Delta A_{CP}$}
\label{sec:sec31}
The amplitude of the SCS decay $\Dz \to \pi^+ \pi^-$ can be decomposed as
\begin{equation}
      A (D^0 \to \pi^+ \pi^-) = 
                \lambda_d \left(A_{\rm Tree} + A_{\rm Peng.}^d \right)
              +
                \lambda_s  A_{\rm Peng.}^s
              +
                \lambda_b  A_{\rm Peng.}^b \; ,
      \label{amplitudeA}
      \end{equation}
with a tree-level amplitude $A_{\rm Tree}$ accompanied by the CKM structure 
$ \lambda_d $ and three penguin contributions $A_{\rm Peng.}^q$
with the internal quark $q=d,s,b$ and the CKM structure
$ \lambda_q $. Using the unitarity of the CKM matrix, all additional, more complicated, contributions such as
re-scattering effects can be included in the same scheme to obtain 
\begin{equation}
A (D^0 \to \pi^+ \pi^-) =  \frac{G_F}{\sqrt{2}} \lambda_d \; { T} 
      \left[ 1 + \frac{\lambda_b}{\lambda_d} { \frac{ P}{T}}\right] \; .
\label{amplitudeB}
\end{equation}
Here $T$ contains not only pure tree-level contributions, but also penguin topologies ($P$), weak exchange ($E$) insertions 
and rescattering ($R$) effects, while $P$ consists of tree-insertion of penguin operators and
 penguin-insertions of tree-level operators:
 \begin{eqnarray}
  T  =  \! \sum \limits_{i = 1,2} C_i \langle Q_i^d \rangle^{T+P+E+R} 
 -  \! \sum \limits_{i = 1,2} C_i \langle Q_i^s \rangle^{P+R}  \! ,  \!
& & \! \! 
P  =  \sum \limits_{i > 3} C_i \langle Q_i^b \rangle^T 
 -  \sum \limits_{i = 1,2}    C_i \langle Q_i^s \rangle^{P+R} \! .
\end{eqnarray}
 Physical observables, for example branching ratios, Br, or direct $\CP$ asymmetries, $a_{\rm dir}$, can be expressed in terms of 
 $|T|$, $|P/T|$ and the strong
phase $\Phi^S = \arg( P/T)$ as
\begin{eqnarray}
\textrm{Br} &\propto & \frac{G_F^2}{2} |\lambda_d|^2 { |T|^2}  \left|  1 + \frac{\lambda_b}{\lambda_d} { \frac{ P}{T}} \right|^2
\approx 
\frac{G_F^2}{2} |\lambda_d|^2 { |T|^2}  
\; ,
\label{BR_approx}
\\
 a_{\rm dir} & = &  \frac{-2 \left|\frac{\lambda_b}{\lambda_d}\right| \sin \gamma \left|\frac{ P}{ T}  \right| \sin \Phi^S}
                                      {1  - 2 \left|\frac{\lambda_b}{\lambda_d}\right| \cos  \gamma  \left|\frac{ P}{ T}  \right| \cos \Phi^S
                                        +  \left|\frac{\lambda_b}{\lambda_d}\right|^2 \left|\frac{ P}{ T}  \right| ^2}
               \approx   -13 \times 10^{-4}
                 \left|\frac{ P}{ T}                \right| \sin \Phi^S \; .
                 \label{CPasym}
\end{eqnarray}
The approximations on the r.h.s  are based on $|\lambda_b/\lambda_d| \approx 7 \times 10^{-4}$.
For the $\Dz \to K^+ K^-$ decay the same formalism applies with obvious replacements, and we find (neglecting mixing-induced $\CP$ violation)
\begin{eqnarray}
\left| \Delta A_{\CP} \right| & \approx & 13 \times 10^{-4} \left|  \left|\frac{ P}{ T}  \right|_{K^+K^-} \sin \Phi^S _{K^+K^-}  + \left|\frac{ P}{ T}  \right|_{\pi^+\pi^-} \sin \Phi^S_{\pi^+\pi^-} \right|
 \,.
\end{eqnarray}
In order to quantify the possible size of the direct $\CP$ violation, we need to know $|P/T|$ and
the strong phase $\Phi^S$ for both decays. 
Since the branching ratios  are well measured~\cite{Zyla:2020zbs},
\begin{eqnarray}
\mathrm{Br}(\Dz \to K^+ K^-)      =   4.08\pm 0.06 \times 10^{-3}, \nonumber
&&
\mathrm{Br}(\Dz \to \pi^+ \pi^-)  =  1.455\pm0.024 \times 10^{-3}, \nonumber
\end{eqnarray}
their values can be used to estimate the size of $T$ via Eq.~(\ref{BR_approx}). With this information, one can then determine an upper SM bound
by maximizing the strong phase and using only a theory estimate for the size of $P$.

\subsubsection{SM  estimates}
The magnitude of $\Delta A_{CP}$ is larger than naive expectations, and it is currently unclear whether this value  
 is governed 
by the SM or whether we are already 
seeing a first glimpse of BSM effects.

Naive perturbative estimates yield $|P/T| \approx 0.1$ (see {\it e.g.} Refs. ~\cite{Grossman:2006jg,Bigi:2011re,Lenz:2013pwa,Grossman:2019xcj}) resulting in the upper bound
\begin{eqnarray}
\left| \Delta A_{CP} \right| & \approx & 
\leq 2.6 \times 10^{-4} \,,
\end{eqnarray}
which is roughly an order-of-magnitude smaller than what is observed. 

Light-Cone Sum Rules (LCSR)
\cite{Balitsky:1989ry} are a QCD-based
method  allowing the determination of hadronic
matrix elements, including
non-perturbative effects. 
This method has been used by the authors of Ref.~\cite{Khodjamirian:2017zdu}
to predict the {\CP} asymmetries in  neutral $D$-meson decays.
Extracting values of the matrix element $|T|$ from
the experimental measurements of the branching ratios, the magnitudes and phases of $P$ are then determined with LCSRs. In this study no prediction is made  of the relative strong phase 
between the tree-level $T$ and penguin $P$ contributions, which therefore  remains a free parameter. 
This calculation has been numerically
updated in  Ref.~\cite{Chala:2019fdb}
and yields:
\begin{eqnarray}
 \left|\frac{ P}{ T}  \right|_{\pi^+\pi^-}   =   0.093 \pm 0.056 \, ,
&&
 \left|\frac{ P}{ T}  \right|_{K^+K^-}   =  0.075 \pm 0.048 \,,
\end{eqnarray} 
which gives a SM bound for $\Delta A_{CP}$ of
\begin{equation}
|\Delta A_{CP}| \le (2.2 \pm 1.4) \times 10^{-4} \le 3.6 \times 10^{-4}.
\label{DeltaACP_estimate}
\end{equation}
It is interesting to note that this result agrees very well with the naive perturbative estimate.
In future one could try to compute both
$T$ and $P$ hadronic matrix elements
entirely with the LCSR method. 
In that case, one
would be able to predict the relative
strong phases and as a consequence arrive at
a more robust SM prediction for 
$\Delta A_{CP}$. Further general
arguments in favour of a small SM
expectation for direct $\CP$ violation in the charm sector may be found in Ref.~\cite{Nierste:2020eqb}.

In contrast the authors of 
Ref.~\cite{Grossman:2019xcj} assert that the magnitude of $\Delta A_{\CP}$ is consistent with having a  SM origin. They employ
$U$-spin and $SU(3)$ flavour symmetry, and conclude that only modest $SU(3)_F$ breaking is required to generate the observed value. This  breaking can come either through perturbative effects or non-perturbative enhancement
due to rescattering. Similar conclusions are reached in Ref.~\cite{Cheng:2019ggx}.

Thus we are in the unfortunate
situation that perturbative and sum-rule estimates are at least one order-of-magnitude  below the experimental
value, while symmetry-based approaches
suggest that the SM is
in perfect agreement with data.
In order to identify the true origin of
direct $\CP$ violation in
the charm sector, greater theoretical understanding is
necessary. Furthermore, new measurements in control channels, discussed  in Sect.~\ref{sec:control}, will be invaluable in confronting the SM predictions arising from the above approaches.

\subsubsection{BSM explanations} 
\label{subsub:BSM}
The possibility that the value of $\Delta A_{\CP}$ is driven by
BSM effects was recently investigated
in Refs.~\cite{Chala:2019fdb,Dery:2019ysp,Bause:2020obd}.
Chala {\it et al.}~\cite{Chala:2019fdb} proposed a $Z'$
model that is severely constrained by $D$-mixing, but not yet ruled out.
Nir {\it et al.}~\cite{Dery:2019ysp}  studied $\Delta A_{\CP}$ in the context of the two Higgs-doublet model (2HDM), minimal supersymmetry (MSSM) and models with vector-like up-quarks, also taking into account $\epsilon'/\epsilon$ bounds from the kaon sector.
Hiller {\it et al.}~\cite{Bause:2020obd} considered more 
general $Z'$ models to explain $\Delta A_{CP}$ and also assessed 
their consequences for rare charm decays. In this study 
the severe constraints from $D$-mixing are
considerably softened by allowing BSM contributions
with different chiralities, which can cancel in the mixing.

\subsubsection{Control measurements}
\label{sec:control}

Further experimental studies are desirable to elucidate the currently unclear 
theory situation, with several decay channels proposed for investigation.

In order to validate the symmetry-based approaches several measurements are desirable.  First of all it is noted that $U$-spin predicts the important relation 
$a_{\rm dir} (K^+ K^-) 
= - 
a_{\rm dir} (\pi^+ \pi^-)
$ (see {\it e.g.} Ref.~\cite{Grossman:2006jg}), which strongly motivates individual determinations of the two asymmetries.
Taking first order $SU(3)_F$
corrections into account one can derive
a sum rule relating 
$a_{\rm dir} (K^+ K^-) $
and
$a_{\rm dir} (\pi^+ \pi^-)$ 
with $a_{\rm dir} (\pi^0 \pi^0)$, the corresponding $\CP$ asymmetry in $\Dz \to \pi^0\pi^0$~\cite{Muller:2015rna}. In 
Ref.~\cite{Cheng:2019ggx} a long list of
predictions for $\CP$ asymmetries is given based on
symmetry considerations; promising modes include $\Dz \to \pi^+ \rho^-$, $\Dz \to K^+ K^{*-}$, $D^+ \to K^+ K^{*0}$, $D^+ \to \eta \rho^+$ , $ D_s^+ \to \pi^+ K^{*0}$ and $D_s^+ \to \pi^0 K^{*+}$.

Another interesting option for control measurements is null tests, {\it i.e.}
decays that are expected to have a
vanishing $\CP$ asymmetry in the SM. 
An example is the SCS mode 
$D^+ \to \pi^+ \pi^0$, which  has the
same leading tree diagram as $D^0 \to
\pi^+  \pi^-$, but receives no contributons from a gluonic-penguin amplitude on account of  the final
state having an isospin value of 2. 
Thus, the SM expectation for  the
$\CP$ asymmetry of this decay is zero~\cite{Brod:2011re}. Null tests involving rare charm decays are discussed in Ref.~\cite{Bause:2019vpr}.

Large $\CP$ violating effects of up to $1\%$ are expected in $\Dz \to K^0_{\rm S}  K^0_{\rm S}$ within the SM, as explained in {\it e.g.} Ref.~\cite{Nierste:2015zra}.
Similarly, the $\CP$ asymmetry in  $\Dz \to K^{*0}  K^0_{\rm S}$ can be as large as $0.3 \%$~\cite{Nierste:2017cua}. 
Additional examples of control channels can be found in Ref.~\cite{Lenz:2013pwa}.

\subsection{Analysis of multi-body decays}

Any direct $\CP$ violation present in a multi-body decay will vary over phase space, and has the possibility to be larger than in the two-body case on account of interference effects between resonances.   For these systems, therefore, $CP$ violation studies must be optimised to search for phase-space-dependent differences in behaviour between the charm and anti-charm decays.  This change in strategy brings some experimental benefits.  For example, tagging and production asymmetries are no longer a primary concern, as they can only induce a global offset in measured rates, although correlations with the trigger and reconstruction of the signal decay may lead to this offset acquiring some non-uniformity at second order. 

LHCb have analysed data sets of between 1.0~and 3.0\,fb$^{-1}$ to study the modes $D^+ \to \pi^+\pi^-\pi^+$~\cite{Aaij:2013jxa}, $\Dz \to \pi^+\pi^-\pi^0$~\cite{Aaij:2014afa}, $\Dz \to K^0_{\rm S}K^\mp\pi^\pm$~\cite{Aaij:2015lsa}, $\Dz \to \pi^+\pi^-\pi^+\pi^-$~\cite{Aaij:2016nki,Aaij:2013swa} and $\Dz \to K^+K^-\pi^+\pi^-$~\cite{Aaij:2018nis,Aaij:2014qwa,Aaij:2013swa}.  The $B$-factories performed important studies of the channels  $D^+ \to K^+K^-\pi^+$~\cite{Lees:2012nn}, $\Dz \to \pi^+\pi^-\pi^0$ and $K^+K^-\pi^0$~\cite{Aubert:2008yd}, $D^+ \to K^+K^0_{\rm S}\pi^+\pi^-$~\cite{Lees:2011dx}, and $\Dz \to K^+K^-\pi^+\pi^-$~\cite{delAmoSanchez:2010xj,Kim:2018mtf}.
All results  are consistent with $\CP$ conservation. No analysis of interesting sensitivity has yet been made of multi-body $D_s^+$ meson decays.  The Run 2 sample of LHCb remains to be exploited for multi-body measurements, and has the potential to bring a great improvement in precision. 

$\CP$-violation searches in phase space can be performed in either a model-dependent or -independent manner.  In the former, an amplitude model is constructed and then fitted separately to the charm and anti-charm decay samples~\cite{Aaij:2015lsa,Aaij:2018nis,Lees:2012nn,Aubert:2008yd}. Significant differences in the amplitude modulus, phase or fit-fraction of any resonance contribution between the two data sets would signify the presence of $\CP$ violation. This approach has the virtue of targeting those regions of phase space where intermediate states are known to exist, and yielding results that are straightforward to interpret. For example, in the $\Dz \to K^+K^-\pi^+\pi^-$ study of LHCb, $\CP$ asymmetries related to the amplitude modulus, phase and fit fraction are determined with uncertainties of 1\% to 15\%, depending on the resonance.

\begin{figure}[htb]
    \centering
    \includegraphics[width=0.85\textwidth]{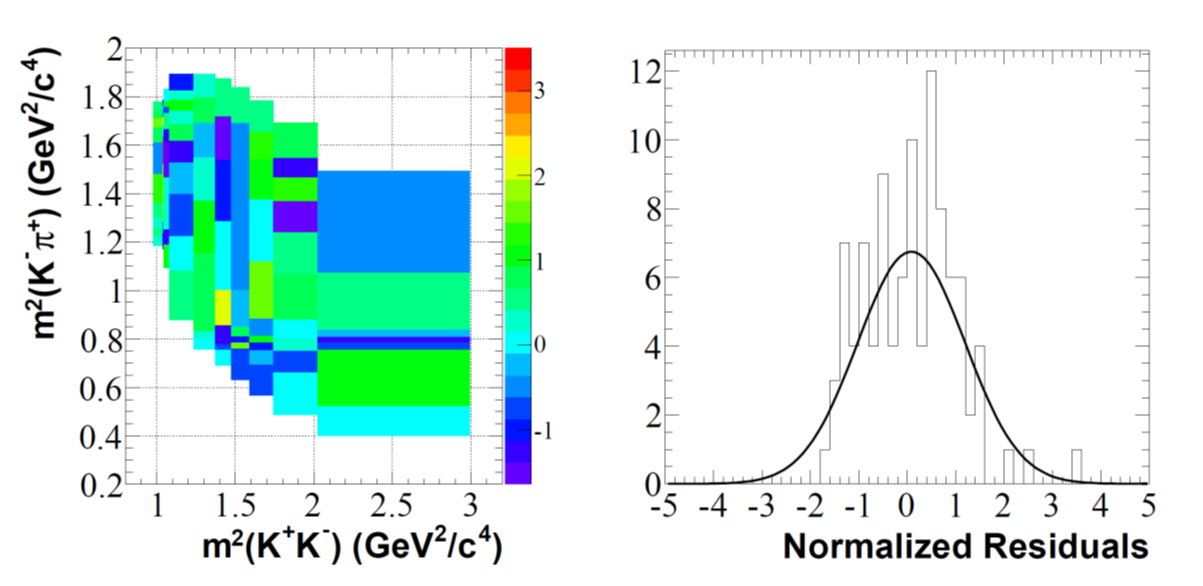}
    \caption{Search for local $\CP$ violation in $D^+ \to K^+K^-\pi^+$ decays. Normalised residuals, expressing the difference between $D^+$ and $D^-$ decays, are determined in suitably chosen bins of the Dalitz plot~\cite{Lees:2012nn}.}
    \label{fig:locaclacp}
\end{figure}

A simple model-independent procedure is to partition phase space into bins, and then search for significant differences in the bin contents between charm and anti-charm decays~\cite{Bediaga:2009tr,Aaij:2013jxa,Aaij:2013swa,Lees:2012nn,Aubert:2008yd}. See, for example, Fig.~\ref{fig:locaclacp}.  Unbinned approaches include the study of angular moments of the intensity distributions~\cite{Lees:2012nn,Aubert:2008yd}, the $k$-nearest neighbour method~\cite{Williams:2010vh,henze1988,schilling,Aaij:2013jxa}, and the so-called energy test~\cite{Williams:2011cd,Aaij:2014afa,Aaij:2016nki}.  When LHCb apply the energy test to the channel $\Dz \to \pi^+\pi^-\pi^+\pi^-$ in the Run 1 data sample, the $P$-odd $\CP$ asymmetries are found to be only marginally consistent with the $\CP$-conservation hypothesis~\cite{Aaij:2016nki}, making this mode of particular interest for future analyses.

An alternative class of model-independent analysis involves triple-product asymmetries, such as $T$-odd moments, which probe $\CP$ violation through differential distributions~\cite{Bigi:2001sg,Durieux:2015zwa}.  This strategy can be regarded as complementary to those based on studies of $\CP$ asymmetries, and has been employed by LHCb and the $B$-factories to study the decays $\Dz \to K^+K^-\pi^+\pi^-$~\cite{Aaij:2014qwa,delAmoSanchez:2010xj,Kim:2018mtf} and  $D^+ \to K^+K^0_{\rm S}\pi^+\pi^-$~\cite{Lees:2011dx}.

\subsection{Searches for $\CP$ violation in charm-baryon decays}
\label{sec:baryons}

$\CP$ violation in baryon decays remains a largely unexplored domain.  Systematic control  is challenging, as most measurements require understanding the different interaction cross-sections of protons and antiprotons. 
LHCb has performed a study analogous to the two-body $\Delta A_{\CP}$ measurement to determine the difference in phase-space-integrated $\CP$ asymmetries between the modes $\Lambda_c^+ \to pK^+K^-$ and $\Lambda_c^+ \to p\pi^+\pi^-$ and achieved a precision of 1\% with Run 1 data~\cite{Aaij:2017xva}.  LHCb has also searched for local $\CP$ asymmetries within the phase-space of the decay $\Xi_c^+ \to pK^-\pi^+$~\cite{Aaij:2020wil}.  In baryon decays it is  possible to probe for $\CP$ violation by measuring the weak asymmetry parameter $\alpha$ in an angular analysis~\cite{Korner:1991ph}, and comparing the results for charm and anti-charm decays.  Until now this method has only been deployed in CF decays~\cite{Hinson:2004pj,Link:2005ft}.

\section{Conclusions and outlook}
\label{sec:conclusions}

The current era is the most exciting in charm physics for many decades.  Neutral-mixing and $\CP$ violation, for long feared to be too small for experimental study, are now observed, and the next goals are firmly in sight.  
The most urgent tasks are to
establish whether the parameter $x$, 
and hence the mass-splitting in the
neutral charm system, is of a similar
magnitude to $y$, or instead vanishing;
to make further measurements of direct
$\CP$ violation, in particular those
that will help elucidate whether the
size of $\Delta A_{\CP}$ is compatible
with SM expectations;  and
finally to intensify the search for
$\CP$ violation associated with
$\Dz\Dzb$ oscillations.  Fortunately,
the prospects for meeting these
challenges are excellent. Many key
measurements are still to be performed
on the Run 1 and Run 2 LHCb data sets, and soon complementary charm studies
will emerge from Belle II.  Even more
excitingly, the vast increase in sample
sizes that will become available at LHCb
Upgrades I and II will allow for a
corresponding advance in precision,
provided that systematic control can be
maintained.  With this likely progress
in mind it is imperative that
theoretical developments keep track. 
Here considerable improvements in
methods based on symmetry principles,
heavy-quark expansion, LCSR sum rules
and lattice QCD are foreseen.

Charm is now a fast-moving discipline, and one that can be considered as complementary to beauty for its potential to test the CKM paradigm and to probe for New-Physics effects.  For flavour physicists this is truly the age of charm.

\section*{DISCLOSURE STATEMENT}
 The authors are not aware of any affiliations, memberships, funding, or financial holdings that
might be perceived as affecting the objectivity of this review. 

\section*{ACKNOWLEDGMENTS}
AL would like to thank Maria Laura Piscopo, Aleksey Rusov  and Christos Vlahos for proof-reading, checking some formulae and for creating Fig.~\ref{fig:loop}, and Marco Gersabeck and Alex Kagan for helpful discussions. GW would like to acknowledge the valuable input of Sneha Malde and Nathan Jurik. 

%

\bibliographystyle{ar-style5}
\bibliography{references}

\end{document}